\UseRawInputEncoding
\documentclass[reprint,showpacs,twocolumn,aps,superscriptaddress]{revtex4-2}
\usepackage{amsfonts}
\usepackage{amsmath}
\usepackage{amssymb}
\usepackage{graphicx}
\usepackage{verbatim}
\usepackage{textcomp}
\usepackage{hyperref}
\usepackage{upgreek}

\providecommand{\U}[1]{\protect\rule{.1in}{.1in}}

\begin{document}

\title{Fully reconfigurable optomechanical add-drop filters}

\author{Yuechen Lei}
\affiliation{Beijing National Laboratory for Condensed Matter Physics, Institute of Physics, Chinese Academy of Sciences, Beijing 100094, P. R. China}
\affiliation{University of Chinese Academy of Sciences, Beijing 100049, P. R. China}
\author{Zhi-Gang Hu}
\affiliation{Beijing National Laboratory for Condensed Matter Physics, Institute of Physics, Chinese Academy of Sciences, Beijing 100094, P. R. China}
\affiliation{University of Chinese Academy of Sciences, Beijing 100049, P. R. China}
\author{Min Wang}
\affiliation{Beijing National Laboratory for Condensed Matter Physics, Institute of Physics, Chinese Academy of Sciences, Beijing 100094, P. R. China}
\affiliation{University of Chinese Academy of Sciences, Beijing 100049, P. R. China}
\author{Yi-Meng Gao}
\affiliation{Beijing National Laboratory for Condensed Matter Physics, Institute of Physics, Chinese Academy of Sciences, Beijing 100094, P. R. China}
\affiliation{University of Chinese Academy of Sciences, Beijing 100049, P. R. China}

\author{Zhanchun Zuo}
\affiliation{Beijing National Laboratory for Condensed Matter Physics, Institute of Physics, Chinese Academy of Sciences, Beijing 100094, P. R. China}
\affiliation{Songshan Lake Materials Laboratory, Dongguan 523808, Guangdong, P. R. China}

\author{Xiulai Xu}
\affiliation{State Key Laboratory for Mesoscopic Physics and Frontiers Science Center for Nano-optoelectronics, School of Physics, Peking University, 100871 Beijing, China}

\author{Bei-Bei Li}
\email{libeibei@iphy.ac.cn}
\affiliation{Beijing National Laboratory for Condensed Matter Physics, Institute of Physics, Chinese Academy of Sciences, Beijing 100094, P. R. China}
\affiliation{Songshan Lake Materials Laboratory, Dongguan 523808, Guangdong, P. R. China}

\date{\today}

\begin{abstract}
Fully reconfigurable add-drop filters (ADFs) have important applications in optical communication and information processing. Here we demonstrate a broadly tunable add-drop filter based on a double-disk cavity optomechanical system, side-coupled with a pair of tapered fiber waveguides. By varying the coupling rates between the cavity and the two waveguides, we investigate the dependence of the through (drop) efficiency on the coupling rates, which agrees well with the theoretical results. By optimizing the cavity-waveguide coupling rates, a drop efficiency of 89$\%$ and a transmission of 1.9$\%$ have been achieved. Benefiting from the large optomechanical coupling coefficient of the double-disk microcavity, a tuning range of 8~nm has been realized, which is more than one free spectral range (FSR) of the cavity. This is realized by changing the air gap of the double disk using a fiber tip, which is controlled by a piezoelectrical nanostage, with a required voltage of 7~V. As a result, both the through and drop signals can be resonant with any wavelength within the transparent window of the cavity material, which indicates that the ADF is fully reconfigurable.
\end{abstract}

\maketitle

\section{Introduction}
Whispering-gallery-mode (WGM) microcavities with high quality ($Q$) factors and small mode volumes can greatly enhance light-matter interactions  \cite{2003N}, and therefore have a wide range of applications in quantum optics \cite{2006N,2021OE}, integrated photonic network \cite{2021S}, optical filters \cite{2002JLT,2009JLT,1998IEEE,1999IEEE2,2002IEEE,2004OE,2012JLT,2016IEEE,2004IEEE,2004PRL,2004IEEE,2013APL,2018IEEE}, cavity optomechanics \cite{2014RMP}, microscale highly sensitive sensors \cite{2020M,2021nanoph}, optical frequency combs \cite{2020S,2022NP}, etc. Among them, the add-drop filter (ADF) is one of the key elements of optical information processing.
The add-drop configuration consisting of a high $Q$ WGM microcavity side-coupled with two waveguides can ideally filter out all the power from the bus waveguide and transfer it to the other drop waveguide, when the input laser is on resonance with the cavity. This represents a crosstalk of zero and a drop efficiency of 100$\%$ \cite{1997JLT}.
 
In order to be resonant with any device of the scalable photonic network, the ADF needs to be fully reconfigurable for applications including optical routers, switches, modulators, and miniaturized spectrometers. This requires the WGM cavities in these ADFs to be tunable over a full free spectral range (FSR). This means the WGM cavity can be resonant with any wavelength within the transparent window of the cavity material \cite{2005IEEE,2008OE2,2011OE,2018OE,2009PRL,2010OL,2018OE2,2013APL2,2017OL,2018AM,2018ACS}. In addition to the large tuning range, high tuning speed and low power consumption are also required for applications of ADFs in integrated photonic networks. FSR tunable microcavities have many other applications, such as cavity quantum electrodynamics \cite{2006N,2021OE} and tunable microlasers \cite{2010APL,2017OE,2016SR,2018OL,2017OL,2018ACS,2019nanoph,2018AM}.

Tuning of the microcavity resonance frequencies is usually realized either by changing the refractive index of the cavity material through thermal-optic effect \cite{2004APL,2005IEEE,2008OE2,2013IEEE,2017OE3}, electro-optic effect \cite{2007OL,2007NP2}, and free carrier injection \cite{2002IEEE2,2006IEEE}, or by altering the round-trip length through applying a stress to the cavity \cite{2009PRL,2010OL,2011OL,2011OE2,2016OE,2017OE2,2018OE2,2020N}. However, achieving a full FSR tuning with high speed has proved challenging using these methods. For instance, thermo-optic tuning requires a large temperature increase, which increases the power consumption and affects the filter characteristic due to thermal expansion. In addition, thermal-optic response is usually quite slow \cite{2004APL,2008OE2,2013IEEE,2017OE3}. Electro-optic tuning and free carrier injection only apply to certain materials, and the realized tuning range are relatively small \cite{2007NP,2007OL,2006IEEE}. Stress tuning to change the cavity circumference usually has a limited tuning range due to the strain limitations of the materials \cite{2009PRL,2017OE2,2018OE2}. In recent years, tuning schemes based on strong optomechanical coupling systems have been demonstrated, including WGM double-disk microavities \cite{2009NP,2009N,2009OE,2018OE,2011OE,2021PRapp} and two coupled photonic crystal cavities \cite{2010APL,2010OE,2011OE3,2012OE,2012NC}. Wide tuning ranges have been realized, by manipulating the mechanical degrees of freedom in these systems. For example, a tuning range over 32~nm (two FSRs) has been realized using optical gradient forces in a silicon nitride double-disk cavity \cite{2011OE} More recently, electrical tuning of 9~nm which is more than three FSRs has been demonstrated in a double-disk cavity with interdigitated electrodes fabricated on the disk \cite{2018OE}.

Here we demonstrate a fully reconfigurable add-drop filter based on a silica double-disk cavity optomechanical system, side-coupled with two tapered fiber waveguides. We study the relation between the through/drop efficiency and the system losses, including the intrinsic decay rate ($\kappa_i$) of the cavity, and the coupling rates with both the bus waveguide ($\kappa_b$) and the drop waveguide ($\kappa_d$). Through optimizing the coupling rates $\kappa_b$ and $\kappa_d$, a drop efficiency as high as 89$\%$ and a through efficiency of 1.9$\%$ has been realized, with a bandwidth of 41.6~GHz for both the through and drop signals. We then tune the cavity resonances by gradually decreasing the air gap of the double disk using a fiber tip. Benefitting from the large optomechanical coupling coefficient (25~GHz/nm) of the double-disk cavity, using an electrical voltage of 7~V has enabled a broad tuning range of around 8~nm, which is more than one FSR of the cavity in 1500~nm. This indicates that the through and drop signals can be resonant with any wavelength in the transparent window of silica, from visible to near infrared.

\begin{figure*}[htb]
\begin{center}
\includegraphics[width=16cm]{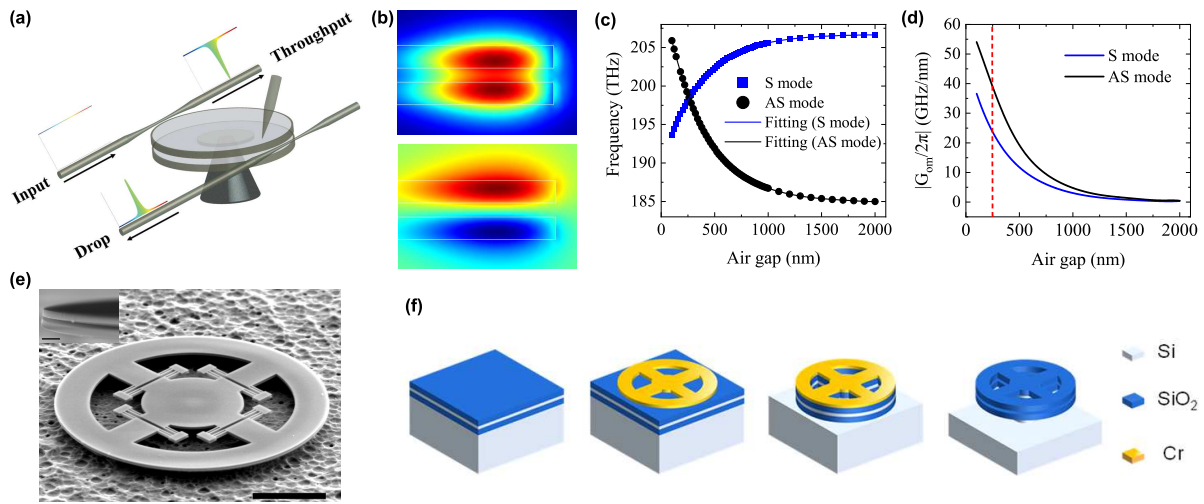}
\end{center}
\caption{(a) Schematic of the tunable add-drop filter. (b) FEM simulation results of the field distribution for the $S$ (upper panel) and $AS$ (lower panel) TE fundamental modes of the double-disk cavity. (c) FEM simulated frequencies of the $S$ and $AS$ modes versus the air gap of the double disk. The azimuthal mode numbers are m$_{S}$=199 and m$_{AS}$=174. (d) Optomechanical coupling coefficients $|{G_{om}/2\pi}|$ of the $S$ and $AS$ modes versus the air gap of the double disk. The dashed line denotes the air gap of the double disk we use in the experiment. (e) SEM image of the double-disk cavity (scale bar of 10 ~\textmu m), with the inset showing details of the air gap of the double disk (scale bar of 1~\textmu m). (f) Schematic of the fabrication processes of the double-disk cavity.}
\label{fig1}
\end{figure*}

\section{Device design and fabrication}

Figure \ref{fig1}(a) shows a schematic diagram of the double-disk add-drop filter. It consists of an on-chip double disk cavity, side-coupled with a pair of tapered fibers, and a fiber tip on the top disk. The double-disk cavity is composed of two thin silica disks with a small air gap between them. When the input laser is on resonance with the cavity, a Lorentzian lineshaped dip (peak) appears in the transmission spectrum at the throughput (drop) port. Therefore the add-drop filter can realize the functions of band-rejection and band-pass at the throughput and drop ports, respectively. The cavity resonance can be tuned through changing the air gap of the double disk, by pressing the top disk with the fiber tip.

Due to the small air gap between the two disks, their optical modes are strongly coupled, producing symmetric ($S$) and antisymmetric ($AS$) modes. An example of the typical optical field distributions of the $S$ (upper panel) and $AS$ (lower panel) modes are shown in Fig. \ref{fig1}(b), obtained by finite element method (FEM) simulation, for a silica double disk with a diameter $d$=80~\textmu m, a thickness $t$=400~nm for each disk, and an air gap $x$=250~nm. It can be seen that, the $S$ ($AS$) mode has a symmetric (antisymmetric) field distribution in the two disks, and a large (zero) field strength in the middle of the gap. It is found that the $Q$ factor of the $S$ mode is generally higher than the $AS$ mode, as the $S$ mode is more confined compared with the $AS$ mode. For instance, the radiation loss dominated $Q$ factors are obtained from the FEM simulation to be $10^{14}$ and $10^4$ for the $S$ and $AS$ modes, respectively.

In order to theoretically analyze the tuning capability of the double-disk cavity, we use FEM to calculate the frequencies of the $S$ and $AS$ fundamental transverse electric (TE) modes, as a function of the air gap ranging from 100~nm to 2~\textmu m. The azimuthal mode numbers of the $S$ and $AS$ modes are $m_{S}$=199 and $m_{AS}$=174. The simulated frequencies for the $S$ and $AS$ modes are shown in the blue squares and black dots in Fig. \ref{fig1}(c), with the blue and black curves showing their polynomial fitting results. It can be seen that, the frequency of the $S$ ($AS$) mode becomes higher (lower) with increasing the air gap. The frequencies shift towards opposite directions for the two modes, which can be understood by the change of the effective refractive index ($n_{eff}$) of the modes \cite{2007NP}. For the $S$ mode, more optical field will be distributed in air with a larger air gap, leading to a smaller $n_{eff}$ (therefore a higher frequency). While for the $AS$ mode, with the increase of the air gap, more field will be distributed in silica, resulting in a larger $n_{eff}$ and a lower frequency.

\begin{figure*}[tbh]
\begin{center}
\includegraphics[width=16cm]{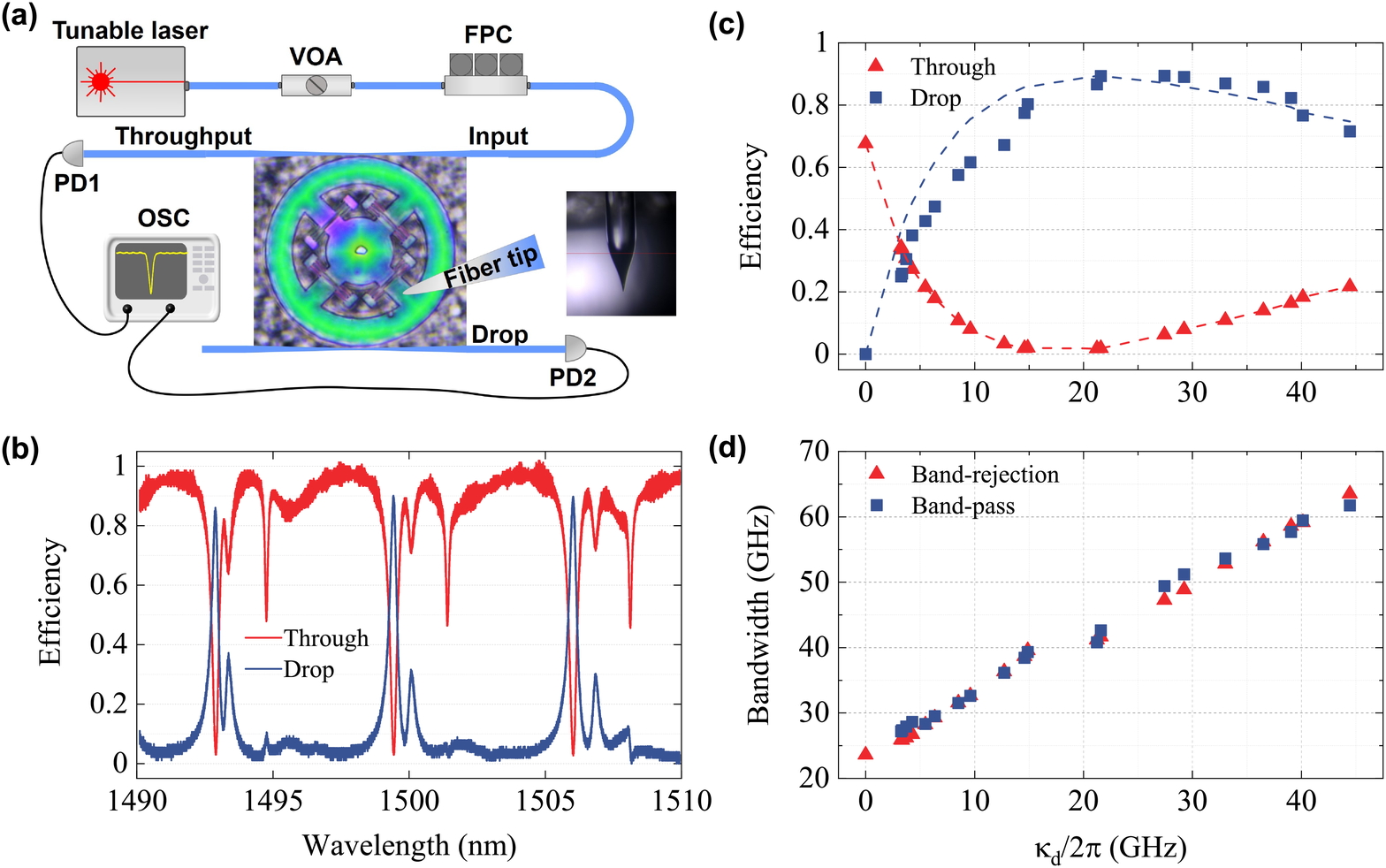}
\end{center}
\caption{(a) Schematic of the experimental measurement setup of the broadly tunable ADF based on the double-disk cavity, with an optical image of the double disk shown in the middle. On the right it shows the optical image of the fiber tip used to press the top disk. VOA: variable optical attenuator, FPC: fiber polarization controller, PD: photodetector, OSC: oscilloscope. (b) Normalized through (red curve) and drop (blue curve) efficiencies. (c) Measured through (red triangles) and drop (blue squares) efficiencies, versus the coupling rate between the cavity and the drop waveguide $\kappa_d$. The red and blue curves are the fitting results using formula (1) and (2). The highest measured drop efficiency is 89$\%$, and the corresponding through efficiency is 1.9$\%$. (d) Measured bandwidths of the band-rejection (red triangles) and band-pass (blue squares) filters, as a function of $\kappa_d$. Both bandwidths are about 41.6~GHz at $\kappa_d/2\pi$=21.5~GHz when it has the highest drop efficiency.}
\label{fig2}
\end{figure*}

By taking a derivative to the fitting curves in Fig. \ref{fig1}(c), we can derive the optomechanical coupling coefficient $|G_{om}|=|d\omega/dx|$ for both the $S$ and $AS$ modes, which quantifies how much the resonance frequency $\omega$ shifts with changing the air gap $x$. Note that the sign of $G_{om}$ is positive (negative) for the $S$ ($AS$) mode. The $|G_{om}/2\pi|$ for the $S$ and $AS$ modes versus the air gap are shown in the blue and black curves of Fig. \ref{fig1}(d). It is worth noting that the $|G_{om}|$ of the $AS$ mode is larger than that of the $S$ mode, which was also obtained in previous work \cite{2009N}. From Fig. \ref{fig1}(d) we can see that, the smaller the air gap is, the larger the optomechanical coupling coefficient is. Therefore, a smaller air gap is desired for a larger tuning range. Taking into account the experimental feasibility, we choose the air gap of the double disk to be 250~nm. This gives $|G_{om}/2\pi|$ of 25~GHz/nm and 39~GHz/nm for the $S$ and $AS$ modes, respectively, which are much larger than that of a single WGM cavity of the same diameter. In addition, the double disk has a larger compliant in the out-of-plane motion than the radial motion in a WGM microdisk, which allows tuning with a smaller force. The FSR for the double disk we use is about 6.2~nm in 1500~nm. Therefore, to achieve a tuning range of more than one FSR, the air gap is required to change by more than 32~nm, which can be easily achieved through pressing the top disk using the fiber tip.

In order to further increase the compliance of the double disk to out-of-plane deformation to facilitate tuning, we design an annulus cavity structure supported by four spokes, as shown in the scanning electron microscope (SEM) image in Fig. \ref{fig1}(e), similar to that used in previous work \cite{2018OE}. The thin tethers between the center disk and the spokes can effectively reduce the buckling effects of the disk originating from the clamping region due to built-in internal stress \cite{2012JMM}. The inset is a close-up on of the double-disk boundary, showing the air gap of around 250~nm. The fabrication process of the device is shown in Fig. \ref{fig1}(f). We first deposit the SiO$_2$/Si/SiO$_2$ stack layers with thicknesses of 400/250/400~nm onto a silicon wafer, using inductively coupled plasma chemical vapor deposition (ICP-CVD) method. The amorphous silicon between the two silica layers is a sacrificial layer which will be etched away to form a silica double disk. In order to reduce the film stress produced in the deposition processes, the wafer is then annealed for 4 hours at 1000~$^{\circ}$C. We then deposit a layer of Chromium (Cr) with a thickness of 70~nm on top of the stack layers through thermal evaporation, and pattern the Cr layer through electron beam lithography (EBL) and lift-off processes, which will be used as the hard mask for etching the stack layers. The SiO$_2$/Si/SiO$_2$ layers are then etched using reactive ion etching (RIE), after which the Cr hard mask is removed. Finally, an isotropic XeF$_2$ etching is performed to etch away both the sacrificial silicon layer and the silicon substrate underneath to suspend the silica double disk.

\section{Measurement and results}


Figure \ref{fig2}(a) shows the schematic of the experimental measurement setup of the broadly tunable ADF. The double-disk cavity evanescently couples with a pair of tapered fibers, with the top (bottom) one serving as the bus (drop) waveguide. An optical microscope image of the double disk is shown in the middle of Fig. \ref{fig2}(a). The distances between the double disk and each tapered fiber is precisely controlled using two 3D piezoelectric nanostages with a precision of 30~nm. Light from a tunable laser in 1500~nm is coupled into the double disk from the input port of the bus waveguide. A variable optical attenuator (VOA) is used to control the input optical power. A fiber polarization controller (FPC) is used to control the polarization of the light to match that of the WGM in the double disk. The transmitted light at the throughput and drop ports are detected by two photodetectors (PDs) separately, and monitored by an oscilloscope (OSC). In order to tune the cavity resonances precisely through changing the air gap of the double disk, we use a fiber tip to press the top disk, with its optical image shown on the right of Fig. \ref{fig2}(a). The position of the fiber tip is controlled by another 3D piezoelectric nanostage. We fabricate the fiber tip through heating and pulling a standard single mode fiber using a CO$_2$ laser and then break it until its diameter reaches about 10~\textmu m. Figure \ref{fig2}(b) shows the through ($T$) and drop ($D$) efficiencies in 1490-1510~nm range, at the throughput and drop ports, defined as the transmitted and dropped powers normalized by the input power. This wavelength range spans three FSRs of the double disk. It can be seen that several modes appear in one FSR, which exhibit resonant dips in the through signal and peaks in the drop signal. 

In order to optimize the drop efficiency, we then study the relation between $D$ and the decay rate of the cavity, including its intrinsic decay rate $\kappa_i$, and its coupling rates $\kappa_b$ and $\kappa_d$ with the bus and drop waveguides. Starting from the equation of motion for the cavity mode, and using the input-output relation, we can obtain the formula \cite{1997JLT,2004PRL,2012JLT,2013APL,2018IEEE} for $T$, $D$, and the relation between them,
\begin{align}
\label{1}
&   T=\frac{4\Delta\omega^{2}+(\kappa_i-\kappa_b+\kappa_d)^{2}}{4\Delta\omega^2+(\kappa_i+\kappa_b+\kappa_d)^2}   \\
&   D=\frac{4\kappa_b\kappa_d}{4\Delta\omega^{2}+(\kappa_i+\kappa_b+\kappa_d)^{2}}    \\
&   D=\frac{1-T}{1+\kappa_i/\kappa_d} 
\end{align} 
where $\Delta\omega$ denotes the detuning between the laser and the cavity resonance frequency. It can be seen that the highest drop efficiency can be obtained when $T=0$ and $\kappa_{i}/\kappa_{d}$ is close to 0. $T=0$ corresponds to a critical coupling condition $\kappa_b=\kappa_i+\kappa_d$ at the resonance frequency. Therefore, in order to achieve the highest drop efficiency in experiments, we can reduce the cavity intrinsic decay rate $\kappa_i$ and keep the coupling rates $\kappa_b$ and $\kappa_d$ as large as possible while maintaining the critical coupling condition. We then optimize the drop efficiency for the mode at 1506~nm which has an intrinsic decay rate $\kappa_i/2\pi$=2.1~GHz. We first only couple the bus waveguide with the double disk and maximize $\kappa_b$ by contacting the bus waveguide on the top disk. We then keep the relative position between the bus waveguide and the double disk, and gradually bring the drop waveguide close to the double disk from the other side. Figure \ref{fig2}(c) shows the measured through (red triangles) and drop (blue squares) efficiencies $T$ and $D$, as a function of $\kappa_d/2\pi$. It can be seen that with the increase of $\kappa_d$, the mode experiences coupling conditions from overcoupling ($\kappa_b>\kappa_d+\kappa_i$) to undercoupling ($\kappa_b<\kappa_d+\kappa_i$). As a result, $T$ decreases first and then increases, with a minimum value $T=1.9\%$ at $\kappa_b/2\pi$=18~GHz and $\kappa_d/2\pi$=21.6~GHz. Correspondingly, the drop efficiency increases first to a maximum value $D=89\%$, and decreases afterwards. The red and blue curves in Fig. \ref{fig2}(c) are the theoretical results derived from formula (1) and (2), with the decay rates ($\kappa_i$, $\kappa_b$ and $\kappa_d$) obtained through Lorentzian fitting of the transmitted and dropped spectra (Fig. \ref{fig2}(b)). It is noted from the fitted linewidths that $\kappa_b$ is slightly changed during the experiment. The slight deviations between the measured and theoretical values of \emph{D} could be from the slightly changed $\kappa_i$ due to the additional scattering loss induced by the drop waveguide. We also study the bandwidths $\kappa=\kappa_i+\kappa_b+\kappa_d$ of the band-rejection and band-pass filters, as a function of $\kappa_d/2\pi$, as shown in the red triangles and blue squares of Fig. \ref{fig2}(d), respectively. As expected, the two bandwidths are approximately the same, and both bandwidths increase with $\kappa_d$. The bandwidth is about 41.6~GHz for the maximum drop efficiency of $89\%$.

\begin{figure}[tb]
\begin{center}
\includegraphics[width=8 cm]{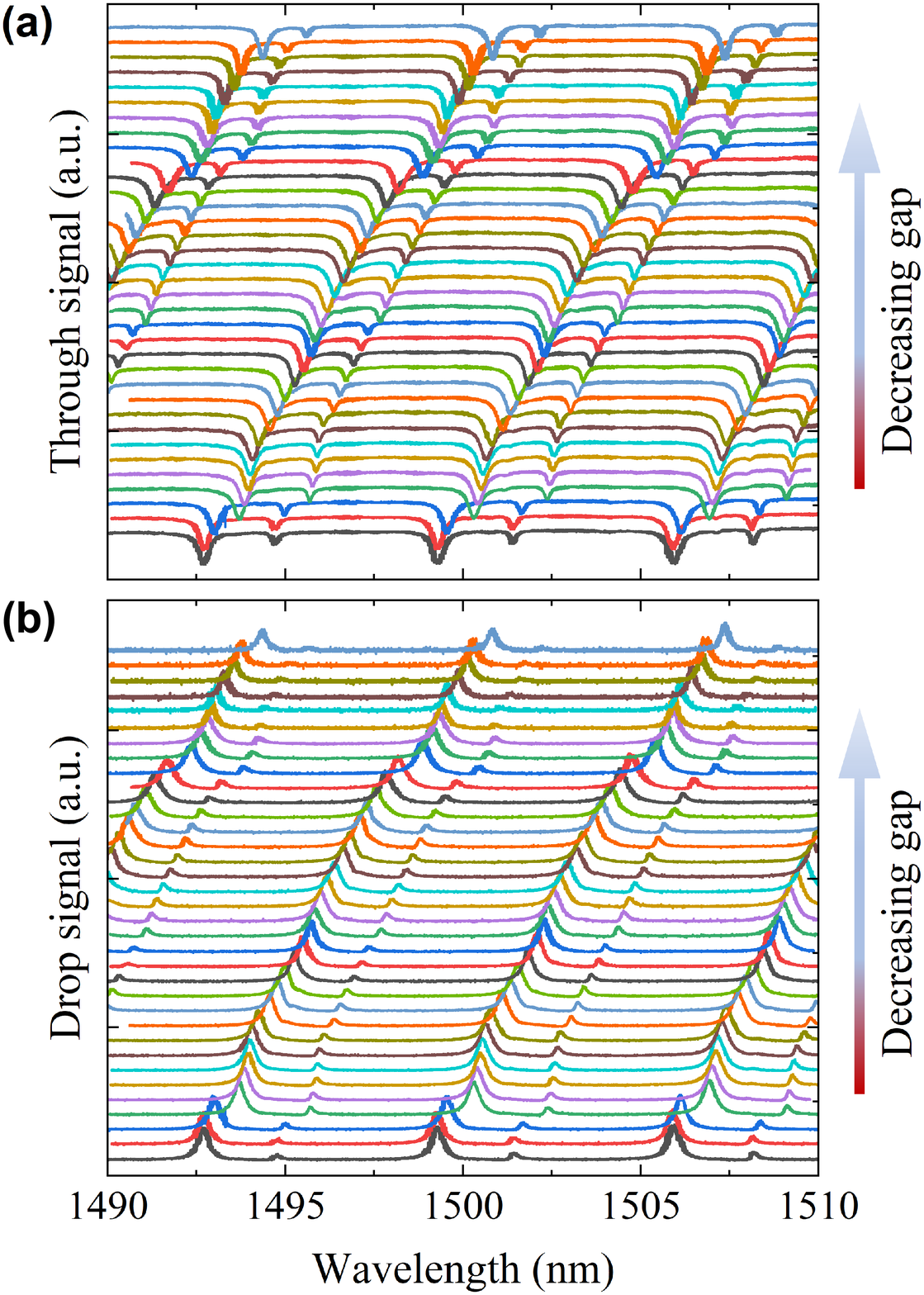}
\end{center}
\caption{(a) and (b) Through and drop signals in 1490-1510~nm range. From the bottom to the top, the air gap is gradually decreased. Both the through and drop signals show a tuning range of around 8~nm which exceeds the FSR of the double-disk cavity.}
\label{fig3}
\end{figure}

In order to realize a broadband tuning of the ADF, we use a fiber tip to press the top disk which can change the air gap of the double disk, while keeping a relatively high drop efficiency. As the position of the fiber tip is gradually lowered by the nanostage, the air gap of the double disk decreases, which shifts the optical resonances to longer (shorter) wavelengths for the $S$ ($AS$) modes. However in our experiment, $AS$ modes are not observed, due to the relatively large loss, which is consistent with the FEM simulation results mentioned above. Figures \ref{fig3}(a) and \ref{fig3}(b) show the through and drop signals in 1490-1510~nm range (spanning 3 FSRs), with decreasing the air gap from the bottom to the top. As the air gap decreases, these modes experience redshift continuously, with a tuning range of around 8~nm, which exceeds one FSR (6.2~nm) of the double-disk cavity. A full FSR tuning means that the through and drop signals can be resonant with any wavelength within the transparent window of cavity material from visible to near infrared, i.e., the ADF is fully reconfigurable. In addition, the tuning speed as high as several MHz can be achieved using this method, which is much faster than that using thermo-optic tuning.  

\section{Summary}
We have demonstrated a fully reconfigurable add-drop filter using a double-disk cavity, side-coupled with a pair of tapered fibers. We investigate the relation between the drop efficiency and the optical decay rates, and achieve a drop efficiency of 89 $\%$ and through efficiency of 1.9$\%$ with band-pass and band-rejection bandwidth of 41.6~GHz. We then realize a broadband tuning of the double disk, through changing the air gap using a fiber tip. Taking advantage of the large optomechanical coupling coefficient of the double-disk cavity, we have achieved a broad tuning range of around 8~nm, which exceeds the FSR of the double-disk cavity. This indicates that filter signals can be resonant with any wavelength within the transparent window of cavity material from visible to near infrared, i.e., the ADF is fully reconfigurable.

The drop efficiency can be further improved by reducing the intrinsic loss of double-disk cavity \cite{2004PRL,2013APL}. The response shape of the filter can be controlled by cascading multiple cavities \cite{2004IEEE,2004OE}. While demonstrated using a silica double disk here, our devices can be transferred to integrated systems, such as silicon nitride \cite{2009N,2011OE} and silicon on insulator \cite{2017SR}. The fully reconfigurable add-drop filters have a range of potential applications in optical communication and information processing, such as on-chip routing \cite{2008OE2,2009NP}, wavelength division multiplexing \cite{1998IEEE} and microscale spectrometers \cite{2021S2}. 

\begin{acknowledgements}

We thank the funding support from The National Key Research and Development Program of China (2021YFA1400700), the National Natural Science Foundation of China (NSFC) (91950118, 12174438, 11934019), and the basic frontier science research program of Chinese Academy of Sciences (ZDBS-LY- JSC003).

Yuechen Lei and Zhi-Gang Hu contributed equally to this work.

\end{acknowledgements}

\bibliography{ref}

\providecommand{\noopsort}[1]{}\providecommand{\singleletter}[1]{#1}%
\begin{thebibliography}{62}%
\makeatletter
\providecommand \@ifxundefined [1]{%
 \@ifx{#1\undefined}
}%
\providecommand \@ifnum [1]{%
 \ifnum #1\expandafter \@firstoftwo
 \else \expandafter \@secondoftwo
 \fi
}%
\providecommand \@ifx [1]{%
 \ifx #1\expandafter \@firstoftwo
 \else \expandafter \@secondoftwo
 \fi
}%
\providecommand \natexlab [1]{#1}%
\providecommand \enquote  [1]{``#1''}%
\providecommand \bibnamefont  [1]{#1}%
\providecommand \bibfnamefont [1]{#1}%
\providecommand \citenamefont [1]{#1}%
\providecommand \href@noop [0]{\@secondoftwo}%
\providecommand \href [0]{\begingroup \@sanitize@url \@href}%
\providecommand \@href[1]{\@@startlink{#1}\@@href}%
\providecommand \@@href[1]{\endgroup#1\@@endlink}%
\providecommand \@sanitize@url [0]{\catcode `\\12\catcode `\$12\catcode
  `\&12\catcode `\#12\catcode `\^12\catcode `\_12\catcode `\%12\relax}%
\providecommand \@@startlink[1]{}%
\providecommand \@@endlink[0]{}%
\providecommand \url  [0]{\begingroup\@sanitize@url \@url }%
\providecommand \@url [1]{\endgroup\@href {#1}{\urlprefix }}%
\providecommand \urlprefix  [0]{URL }%
\providecommand \Eprint [0]{\href }%
\providecommand \doibase [0]{https://doi.org/}%
\providecommand \selectlanguage [0]{\@gobble}%
\providecommand \bibinfo  [0]{\@secondoftwo}%
\providecommand \bibfield  [0]{\@secondoftwo}%
\providecommand \translation [1]{[#1]}%
\providecommand \BibitemOpen [0]{}%
\providecommand \bibitemStop [0]{}%
\providecommand \bibitemNoStop [0]{.\EOS\space}%
\providecommand \EOS [0]{\spacefactor3000\relax}%
\providecommand \BibitemShut  [1]{\csname bibitem#1\endcsname}%
\let\auto@bib@innerbib\@empty
\bibitem [{\citenamefont {Vahala}(2003)}]{2003N}%
  \BibitemOpen
  \bibfield  {author} {\bibinfo {author} {\bibfnamefont {K.~J.}\ \bibnamefont
  {Vahala}},\ }\bibfield  {title} {\bibinfo {title} {Optical microcavities},\
  }\href@noop {} {\bibfield  {journal} {\bibinfo  {journal} {Nature}\ }\textbf
  {\bibinfo {volume} {424}},\ \bibinfo {pages} {839} (\bibinfo {year}
  {2003})}\BibitemShut {NoStop}%
\bibitem [{\citenamefont {Aoki}\ \emph {et~al.}(2006)\citenamefont {Aoki},
  \citenamefont {Dayan}, \citenamefont {Wilcut}, \citenamefont {Bowen},
  \citenamefont {Parkins}, \citenamefont {Kippenberg}, \citenamefont {Vahala},\
  and\ \citenamefont {Kimble}}]{2006N}%
  \BibitemOpen
  \bibfield  {author} {\bibinfo {author} {\bibfnamefont {T.}~\bibnamefont
  {Aoki}}, \bibinfo {author} {\bibfnamefont {B.}~\bibnamefont {Dayan}},
  \bibinfo {author} {\bibfnamefont {E.}~\bibnamefont {Wilcut}}, \bibinfo
  {author} {\bibfnamefont {W.~P.}\ \bibnamefont {Bowen}}, \bibinfo {author}
  {\bibfnamefont {A.~S.}\ \bibnamefont {Parkins}}, \bibinfo {author}
  {\bibfnamefont {T.~J.}\ \bibnamefont {Kippenberg}}, \bibinfo {author}
  {\bibfnamefont {K.~J.}\ \bibnamefont {Vahala}},\ and\ \bibinfo {author}
  {\bibfnamefont {H.~J.}\ \bibnamefont {Kimble}},\ }\bibfield  {title}
  {\bibinfo {title} {Observation of strong coupling between one atom and a
  monolithic microresonator},\ }\href {https://doi.org/10.1038/nature05147}
  {\bibfield  {journal} {\bibinfo  {journal} {Nature}\ }\textbf {\bibinfo
  {volume} {443}},\ \bibinfo {pages} {671} (\bibinfo {year}
  {2006})}\BibitemShut {NoStop}%
\bibitem [{\citenamefont {Yang}\ \emph
  {et~al.}(2021{\natexlab{a}})\citenamefont {Yang}, \citenamefont {Shi},
  \citenamefont {Xie}, \citenamefont {Wu}, \citenamefont {Xiao}, \citenamefont
  {Song}, \citenamefont {Dang}, \citenamefont {Sun}, \citenamefont {Yang},
  \citenamefont {wang}, \citenamefont {Ge}, \citenamefont {Li}, \citenamefont
  {Zuo}, \citenamefont {Jin},\ and\ \citenamefont {Xu}}]{2021OE}%
  \BibitemOpen
  \bibfield  {author} {\bibinfo {author} {\bibfnamefont {J.}~\bibnamefont
  {Yang}}, \bibinfo {author} {\bibfnamefont {S.}~\bibnamefont {Shi}}, \bibinfo
  {author} {\bibfnamefont {X.}~\bibnamefont {Xie}}, \bibinfo {author}
  {\bibfnamefont {S.}~\bibnamefont {Wu}}, \bibinfo {author} {\bibfnamefont
  {S.}~\bibnamefont {Xiao}}, \bibinfo {author} {\bibfnamefont {F.}~\bibnamefont
  {Song}}, \bibinfo {author} {\bibfnamefont {J.}~\bibnamefont {Dang}}, \bibinfo
  {author} {\bibfnamefont {S.}~\bibnamefont {Sun}}, \bibinfo {author}
  {\bibfnamefont {L.}~\bibnamefont {Yang}}, \bibinfo {author} {\bibfnamefont
  {Y.}~\bibnamefont {wang}}, \bibinfo {author} {\bibfnamefont {Z.-Y.}\
  \bibnamefont {Ge}}, \bibinfo {author} {\bibfnamefont {B.-B.}\ \bibnamefont
  {Li}}, \bibinfo {author} {\bibfnamefont {Z.}~\bibnamefont {Zuo}}, \bibinfo
  {author} {\bibfnamefont {K.}~\bibnamefont {Jin}},\ and\ \bibinfo {author}
  {\bibfnamefont {X.}~\bibnamefont {Xu}},\ }\bibfield  {title} {\bibinfo
  {title} {Enhanced emission from a single quantum dot in a microdisk at a
  deterministic diabolical point},\ }\href {https://doi.org/10.1364/OE.419740}
  {\bibfield  {journal} {\bibinfo  {journal} {Opt. Express}\ }\textbf {\bibinfo
  {volume} {29}},\ \bibinfo {pages} {14231} (\bibinfo {year}
  {2021}{\natexlab{a}})}\BibitemShut {NoStop}%
\bibitem [{\citenamefont {Xiang}\ \emph {et~al.}(2021)\citenamefont {Xiang},
  \citenamefont {Liu}, \citenamefont {Guo}, \citenamefont {Chang},
  \citenamefont {Wang}, \citenamefont {Weng}, \citenamefont {Peters},
  \citenamefont {Xie}, \citenamefont {Zhang}, \citenamefont {Riemensberger},
  \citenamefont {Selvidge}, \citenamefont {Kippenberg},\ and\ \citenamefont
  {Bowers}}]{2021S}%
  \BibitemOpen
  \bibfield  {author} {\bibinfo {author} {\bibfnamefont {C.}~\bibnamefont
  {Xiang}}, \bibinfo {author} {\bibfnamefont {J.}~\bibnamefont {Liu}}, \bibinfo
  {author} {\bibfnamefont {J.}~\bibnamefont {Guo}}, \bibinfo {author}
  {\bibfnamefont {L.}~\bibnamefont {Chang}}, \bibinfo {author} {\bibfnamefont
  {R.~N.}\ \bibnamefont {Wang}}, \bibinfo {author} {\bibfnamefont
  {W.}~\bibnamefont {Weng}}, \bibinfo {author} {\bibfnamefont {J.}~\bibnamefont
  {Peters}}, \bibinfo {author} {\bibfnamefont {W.}~\bibnamefont {Xie}},
  \bibinfo {author} {\bibfnamefont {Z.}~\bibnamefont {Zhang}}, \bibinfo
  {author} {\bibfnamefont {J.}~\bibnamefont {Riemensberger}}, \bibinfo {author}
  {\bibfnamefont {J.}~\bibnamefont {Selvidge}}, \bibinfo {author}
  {\bibfnamefont {T.~J.}\ \bibnamefont {Kippenberg}},\ and\ \bibinfo {author}
  {\bibfnamefont {J.~E.}\ \bibnamefont {Bowers}},\ }\bibfield  {title}
  {\bibinfo {title} {Laser soliton microcombs heterogeneously integrated on
  silicon},\ }\href {https://doi.org/10.1126/science.abh2076} {\bibfield
  {journal} {\bibinfo  {journal} {Science}\ }\textbf {\bibinfo {volume}
  {373}},\ \bibinfo {pages} {99} (\bibinfo {year} {2021})}\BibitemShut
  {NoStop}%
\bibitem [{\citenamefont {Grover}\ \emph {et~al.}(2002)\citenamefont {Grover},
  \citenamefont {Van}, \citenamefont {Ibrahim}, \citenamefont {Absil},
  \citenamefont {Calhoun}, \citenamefont {Johnson}, \citenamefont
  {Hryniewicz},\ and\ \citenamefont {Ho}}]{2002JLT}%
  \BibitemOpen
  \bibfield  {author} {\bibinfo {author} {\bibfnamefont {R.}~\bibnamefont
  {Grover}}, \bibinfo {author} {\bibfnamefont {V.}~\bibnamefont {Van}},
  \bibinfo {author} {\bibfnamefont {T.}~\bibnamefont {Ibrahim}}, \bibinfo
  {author} {\bibfnamefont {P.}~\bibnamefont {Absil}}, \bibinfo {author}
  {\bibfnamefont {L.}~\bibnamefont {Calhoun}}, \bibinfo {author} {\bibfnamefont
  {F.}~\bibnamefont {Johnson}}, \bibinfo {author} {\bibfnamefont
  {J.}~\bibnamefont {Hryniewicz}},\ and\ \bibinfo {author} {\bibfnamefont
  {P.-T.}\ \bibnamefont {Ho}},\ }\bibfield  {title} {\bibinfo {title}
  {Parallel-cascaded semiconductor microring resonators for high-order and
  wide-{FSR} filters},\ }\href {https://doi.org/10.1109/JLT.2002.1007947}
  {\bibfield  {journal} {\bibinfo  {journal} {J. Light. Technol.}\ }\textbf
  {\bibinfo {volume} {20}},\ \bibinfo {pages} {900} (\bibinfo {year}
  {2002})}\BibitemShut {NoStop}%
\bibitem [{\citenamefont {Lee}\ \emph {et~al.}(2009)\citenamefont {Lee},
  \citenamefont {Biberman}, \citenamefont {Sherwood-Droz}, \citenamefont
  {Poitras}, \citenamefont {Lipson},\ and\ \citenamefont {Bergman}}]{2009JLT}%
  \BibitemOpen
  \bibfield  {author} {\bibinfo {author} {\bibfnamefont {B.~G.}\ \bibnamefont
  {Lee}}, \bibinfo {author} {\bibfnamefont {A.}~\bibnamefont {Biberman}},
  \bibinfo {author} {\bibfnamefont {N.}~\bibnamefont {Sherwood-Droz}}, \bibinfo
  {author} {\bibfnamefont {C.~B.}\ \bibnamefont {Poitras}}, \bibinfo {author}
  {\bibfnamefont {M.}~\bibnamefont {Lipson}},\ and\ \bibinfo {author}
  {\bibfnamefont {K.}~\bibnamefont {Bergman}},\ }\bibfield  {title} {\bibinfo
  {title} {High-speed 2{\texttimes}2 switch for multiwavelength
  silicon-photonic networks–on-chip},\ }\href
  {https://doi.org/10.1109/JLT.2009.2019256} {\bibfield  {journal} {\bibinfo
  {journal} {J. Light. Technol.}\ }\textbf {\bibinfo {volume} {27}},\ \bibinfo
  {pages} {2900} (\bibinfo {year} {2009})}\BibitemShut {NoStop}%
\bibitem [{\citenamefont {Soref}\ and\ \citenamefont
  {Little}(1998)}]{1998IEEE}%
  \BibitemOpen
  \bibfield  {author} {\bibinfo {author} {\bibfnamefont {R.}~\bibnamefont
  {Soref}}\ and\ \bibinfo {author} {\bibfnamefont {B.}~\bibnamefont {Little}},\
  }\bibfield  {title} {\bibinfo {title} {Proposed {N}-wavelength {M}-fiber
  {WDM} crossconnect switch using active microring resonators},\ }\href
  {https://doi.org/10.1109/68.701522} {\bibfield  {journal} {\bibinfo
  {journal} {IEEE Photon. Technol. Lett.}\ }\textbf {\bibinfo {volume} {10}},\
  \bibinfo {pages} {1121} (\bibinfo {year} {1998})}\BibitemShut {NoStop}%
\bibitem [{\citenamefont {Cai}\ \emph {et~al.}(1999)\citenamefont {Cai},
  \citenamefont {Hunziker},\ and\ \citenamefont {Vahala}}]{1999IEEE2}%
  \BibitemOpen
  \bibfield  {author} {\bibinfo {author} {\bibfnamefont {M.}~\bibnamefont
  {Cai}}, \bibinfo {author} {\bibfnamefont {G.}~\bibnamefont {Hunziker}},\ and\
  \bibinfo {author} {\bibfnamefont {K.}~\bibnamefont {Vahala}},\ }\bibfield
  {title} {\bibinfo {title} {Fiber-optic add-drop device based on a silica
  microsphere-whispering gallery mode system},\ }\href
  {https://doi.org/10.1109/68.766785} {\bibfield  {journal} {\bibinfo
  {journal} {IEEE Photon. Technol. Lett.}\ }\textbf {\bibinfo {volume} {11}},\
  \bibinfo {pages} {686} (\bibinfo {year} {1999})}\BibitemShut {NoStop}%
\bibitem [{\citenamefont {Yanagase}\ \emph {et~al.}(2002)\citenamefont
  {Yanagase}, \citenamefont {Suzuki}, \citenamefont {Kokubun},\ and\
  \citenamefont {Chu}}]{2002IEEE}%
  \BibitemOpen
  \bibfield  {author} {\bibinfo {author} {\bibfnamefont {Y.}~\bibnamefont
  {Yanagase}}, \bibinfo {author} {\bibfnamefont {S.}~\bibnamefont {Suzuki}},
  \bibinfo {author} {\bibfnamefont {Y.}~\bibnamefont {Kokubun}},\ and\ \bibinfo
  {author} {\bibfnamefont {S.~T.}\ \bibnamefont {Chu}},\ }\bibfield  {title}
  {\bibinfo {title} {Box-like filter response and expansion of {FSR} by a
  vertically triple coupled microring resonator filter},\ }\href
  {https://doi.org/10.1109/JLT.2002.800296} {\bibfield  {journal} {\bibinfo
  {journal} {J. Light. Technol.}\ }\textbf {\bibinfo {volume} {20}},\ \bibinfo
  {pages} {1525} (\bibinfo {year} {2002})}\BibitemShut {NoStop}%
\bibitem [{\citenamefont {Barwicz}\ \emph {et~al.}(2004)\citenamefont
  {Barwicz}, \citenamefont {Popovic}, \citenamefont {Rakich}, \citenamefont
  {Watts}, \citenamefont {Haus}, \citenamefont {Ippen},\ and\ \citenamefont
  {Smith}}]{2004OE}%
  \BibitemOpen
  \bibfield  {author} {\bibinfo {author} {\bibfnamefont {T.}~\bibnamefont
  {Barwicz}}, \bibinfo {author} {\bibfnamefont {M.~A.}\ \bibnamefont
  {Popovic}}, \bibinfo {author} {\bibfnamefont {P.~T.}\ \bibnamefont {Rakich}},
  \bibinfo {author} {\bibfnamefont {M.~R.}\ \bibnamefont {Watts}}, \bibinfo
  {author} {\bibfnamefont {H.~A.}\ \bibnamefont {Haus}}, \bibinfo {author}
  {\bibfnamefont {E.~P.}\ \bibnamefont {Ippen}},\ and\ \bibinfo {author}
  {\bibfnamefont {H.~I.}\ \bibnamefont {Smith}},\ }\bibfield  {title} {\bibinfo
  {title} {Microring-resonator-based add-drop filters in {S}i{N}: fabrication
  and analysis},\ }\href {https://doi.org/10.1364/OPEX.12.001437} {\bibfield
  {journal} {\bibinfo  {journal} {Opt. Express}\ }\textbf {\bibinfo {volume}
  {12}},\ \bibinfo {pages} {1437} (\bibinfo {year} {2004})}\BibitemShut
  {NoStop}%
\bibitem [{\citenamefont {Monifi}\ \emph {et~al.}(2012)\citenamefont {Monifi},
  \citenamefont {Friedlein}, \citenamefont {Ozdemir},\ and\ \citenamefont
  {Yang}}]{2012JLT}%
  \BibitemOpen
  \bibfield  {author} {\bibinfo {author} {\bibfnamefont {F.}~\bibnamefont
  {Monifi}}, \bibinfo {author} {\bibfnamefont {J.}~\bibnamefont {Friedlein}},
  \bibinfo {author} {\bibfnamefont {Å.~K.}\ \bibnamefont {Ozdemir}},\ and\
  \bibinfo {author} {\bibfnamefont {L.}~\bibnamefont {Yang}},\ }\bibfield
  {title} {\bibinfo {title} {A robust and tunable add–drop filter using
  whispering gallery mode microtoroid resonator},\ }\href
  {https://doi.org/10.1109/JLT.2012.2214026} {\bibfield  {journal} {\bibinfo
  {journal} {J. Light. Technol.}\ }\textbf {\bibinfo {volume} {30}},\ \bibinfo
  {pages} {3306} (\bibinfo {year} {2012})}\BibitemShut {NoStop}%
\bibitem [{\citenamefont {Wang}\ \emph {et~al.}(2016)\citenamefont {Wang},
  \citenamefont {Madugani}, \citenamefont {Zhao}, \citenamefont {Yang},
  \citenamefont {Ward}, \citenamefont {Yang}, \citenamefont {Farrell},
  \citenamefont {Brambilla},\ and\ \citenamefont {N.~Chormaic}}]{2016IEEE}%
  \BibitemOpen
  \bibfield  {author} {\bibinfo {author} {\bibfnamefont {P.}~\bibnamefont
  {Wang}}, \bibinfo {author} {\bibfnamefont {R.}~\bibnamefont {Madugani}},
  \bibinfo {author} {\bibfnamefont {H.}~\bibnamefont {Zhao}}, \bibinfo {author}
  {\bibfnamefont {W.}~\bibnamefont {Yang}}, \bibinfo {author} {\bibfnamefont
  {J.~M.}\ \bibnamefont {Ward}}, \bibinfo {author} {\bibfnamefont
  {Y.}~\bibnamefont {Yang}}, \bibinfo {author} {\bibfnamefont {G.}~\bibnamefont
  {Farrell}}, \bibinfo {author} {\bibfnamefont {G.}~\bibnamefont {Brambilla}},\
  and\ \bibinfo {author} {\bibfnamefont {S.}~\bibnamefont {N.~Chormaic}},\
  }\bibfield  {title} {\bibinfo {title} {Packaged optical add-drop filter based
  on an optical microfiber coupler and a microsphere},\ }\href
  {https://doi.org/10.1109/LPT.2016.2591959} {\bibfield  {journal} {\bibinfo
  {journal} {IEEE Photon. Technol. Lett.}\ }\textbf {\bibinfo {volume} {28}},\
  \bibinfo {pages} {2277} (\bibinfo {year} {2016})}\BibitemShut {NoStop}%
\bibitem [{\citenamefont {Little}\ \emph {et~al.}(2004)\citenamefont {Little},
  \citenamefont {Chu}, \citenamefont {Absil}, \citenamefont {Hryniewicz},
  \citenamefont {Johnson}, \citenamefont {Seiferth}, \citenamefont {Gill},
  \citenamefont {Van}, \citenamefont {King},\ and\ \citenamefont
  {Trakalo}}]{2004IEEE}%
  \BibitemOpen
  \bibfield  {author} {\bibinfo {author} {\bibfnamefont {B.}~\bibnamefont
  {Little}}, \bibinfo {author} {\bibfnamefont {S.}~\bibnamefont {Chu}},
  \bibinfo {author} {\bibfnamefont {P.}~\bibnamefont {Absil}}, \bibinfo
  {author} {\bibfnamefont {J.}~\bibnamefont {Hryniewicz}}, \bibinfo {author}
  {\bibfnamefont {F.}~\bibnamefont {Johnson}}, \bibinfo {author} {\bibfnamefont
  {F.}~\bibnamefont {Seiferth}}, \bibinfo {author} {\bibfnamefont
  {D.}~\bibnamefont {Gill}}, \bibinfo {author} {\bibfnamefont {V.}~\bibnamefont
  {Van}}, \bibinfo {author} {\bibfnamefont {O.}~\bibnamefont {King}},\ and\
  \bibinfo {author} {\bibfnamefont {M.}~\bibnamefont {Trakalo}},\ }\bibfield
  {title} {\bibinfo {title} {Very high-order microring resonator filters for
  {WDM} applications},\ }\href {https://doi.org/10.1109/LPT.2004.834525}
  {\bibfield  {journal} {\bibinfo  {journal} {IEEE Photon. Technol. Lett.}\
  }\textbf {\bibinfo {volume} {16}},\ \bibinfo {pages} {2263} (\bibinfo {year}
  {2004})}\BibitemShut {NoStop}%
\bibitem [{\citenamefont {Rokhsari}\ and\ \citenamefont
  {Vahala}(2004)}]{2004PRL}%
  \BibitemOpen
  \bibfield  {author} {\bibinfo {author} {\bibfnamefont {H.}~\bibnamefont
  {Rokhsari}}\ and\ \bibinfo {author} {\bibfnamefont {K.~J.}\ \bibnamefont
  {Vahala}},\ }\bibfield  {title} {\bibinfo {title} {Ultralow loss, high {$Q$},
  four port resonant couplers for quantum optics and photonics},\ }\href
  {https://doi.org/10.1103/PhysRevLett.92.253905} {\bibfield  {journal}
  {\bibinfo  {journal} {Phys. Rev. Lett.}\ }\textbf {\bibinfo {volume} {92}},\
  \bibinfo {pages} {253905} (\bibinfo {year} {2004})}\BibitemShut {NoStop}%
\bibitem [{\citenamefont {Monifi}\ \emph {et~al.}(2013)\citenamefont {Monifi},
  \citenamefont {Kaya~Özdemir},\ and\ \citenamefont {Yang}}]{2013APL}%
  \BibitemOpen
  \bibfield  {author} {\bibinfo {author} {\bibfnamefont {F.}~\bibnamefont
  {Monifi}}, \bibinfo {author} {\bibfnamefont {Å.}~\bibnamefont
  {Kaya~Özdemir}},\ and\ \bibinfo {author} {\bibfnamefont {L.}~\bibnamefont
  {Yang}},\ }\bibfield  {title} {\bibinfo {title} {Tunable add-drop filter
  using an active whispering gallery mode microcavity},\ }\href
  {https://doi.org/10.1063/1.4827637} {\bibfield  {journal} {\bibinfo
  {journal} {Appl. Phys. Lett.}\ }\textbf {\bibinfo {volume} {103}},\ \bibinfo
  {pages} {181103} (\bibinfo {year} {2013})}\BibitemShut {NoStop}%
\bibitem [{\citenamefont {Zhou}\ \emph {et~al.}(2018)\citenamefont {Zhou},
  \citenamefont {Chen}, \citenamefont {Shen}, \citenamefont {Zou},
  \citenamefont {Guo},\ and\ \citenamefont {Dong}}]{2018IEEE}%
  \BibitemOpen
  \bibfield  {author} {\bibinfo {author} {\bibfnamefont {Z.-H.}\ \bibnamefont
  {Zhou}}, \bibinfo {author} {\bibfnamefont {Y.}~\bibnamefont {Chen}}, \bibinfo
  {author} {\bibfnamefont {Z.}~\bibnamefont {Shen}}, \bibinfo {author}
  {\bibfnamefont {C.-L.}\ \bibnamefont {Zou}}, \bibinfo {author} {\bibfnamefont
  {G.-C.}\ \bibnamefont {Guo}},\ and\ \bibinfo {author} {\bibfnamefont {C.-H.}\
  \bibnamefont {Dong}},\ }\bibfield  {title} {\bibinfo {title} {Tunable
  add–drop filter with hollow bottlelike microresonators},\ }\href
  {https://doi.org/10.1109/JPHOT.2017.2764071} {\bibfield  {journal} {\bibinfo
  {journal} {IEEE Photon. J.}\ }\textbf {\bibinfo {volume} {10}},\ \bibinfo
  {pages} {1} (\bibinfo {year} {2018})}\BibitemShut {NoStop}%
\bibitem [{\citenamefont {Aspelmeyer}\ \emph {et~al.}(2014)\citenamefont
  {Aspelmeyer}, \citenamefont {Kippenberg},\ and\ \citenamefont
  {Marquardt}}]{2014RMP}%
  \BibitemOpen
  \bibfield  {author} {\bibinfo {author} {\bibfnamefont {M.}~\bibnamefont
  {Aspelmeyer}}, \bibinfo {author} {\bibfnamefont {T.~J.}\ \bibnamefont
  {Kippenberg}},\ and\ \bibinfo {author} {\bibfnamefont {F.}~\bibnamefont
  {Marquardt}},\ }\bibfield  {title} {\bibinfo {title} {Cavity optomechanics},\
  }\href {https://doi.org/10.1103/RevModPhys.86.1391} {\bibfield  {journal}
  {\bibinfo  {journal} {Rev. Mod. Phys.}\ }\textbf {\bibinfo {volume} {86}},\
  \bibinfo {pages} {1391} (\bibinfo {year} {2014})}\BibitemShut {NoStop}%
\bibitem [{\citenamefont {Jiang}\ \emph {et~al.}(2020)\citenamefont {Jiang},
  \citenamefont {Qavi}, \citenamefont {Huang},\ and\ \citenamefont
  {Yang}}]{2020M}%
  \BibitemOpen
  \bibfield  {author} {\bibinfo {author} {\bibfnamefont {X.}~\bibnamefont
  {Jiang}}, \bibinfo {author} {\bibfnamefont {A.~J.}\ \bibnamefont {Qavi}},
  \bibinfo {author} {\bibfnamefont {S.~H.}\ \bibnamefont {Huang}},\ and\
  \bibinfo {author} {\bibfnamefont {L.}~\bibnamefont {Yang}},\ }\bibfield
  {title} {\bibinfo {title} {Whispering-gallery sensors},\ }\href
  {https://doi.org/10.1016/j.matt.2020.07.008} {\bibfield  {journal} {\bibinfo
  {journal} {Matter}\ }\textbf {\bibinfo {volume} {3}},\ \bibinfo {pages} {371}
  (\bibinfo {year} {2020})}\BibitemShut {NoStop}%
\bibitem [{\citenamefont {Li}\ \emph {et~al.}(2021)\citenamefont {Li},
  \citenamefont {Ou}, \citenamefont {Lei},\ and\ \citenamefont
  {Liu}}]{2021nanoph}%
  \BibitemOpen
  \bibfield  {author} {\bibinfo {author} {\bibfnamefont {B.-B.}\ \bibnamefont
  {Li}}, \bibinfo {author} {\bibfnamefont {L.}~\bibnamefont {Ou}}, \bibinfo
  {author} {\bibfnamefont {Y.}~\bibnamefont {Lei}},\ and\ \bibinfo {author}
  {\bibfnamefont {Y.-C.}\ \bibnamefont {Liu}},\ }\bibfield  {title} {\bibinfo
  {title} {Cavity optomechanical sensing},\ }\href
  {https://doi.org/doi:10.1515/nanoph-2021-0256} {\bibfield  {journal}
  {\bibinfo  {journal} {Nanophotonics}\ }\textbf {\bibinfo {volume} {10}},\
  \bibinfo {pages} {2799} (\bibinfo {year} {2021})}\BibitemShut {NoStop}%
\bibitem [{\citenamefont {Diddams}\ \emph {et~al.}(2020)\citenamefont
  {Diddams}, \citenamefont {Vahala},\ and\ \citenamefont {Udem}}]{2020S}%
  \BibitemOpen
  \bibfield  {author} {\bibinfo {author} {\bibfnamefont {S.~A.}\ \bibnamefont
  {Diddams}}, \bibinfo {author} {\bibfnamefont {K.}~\bibnamefont {Vahala}},\
  and\ \bibinfo {author} {\bibfnamefont {T.}~\bibnamefont {Udem}},\ }\bibfield
  {title} {\bibinfo {title} {Optical frequency combs: Coherently uniting the
  electromagnetic spectrum},\ }\href {https://doi.org/10.1126/science.aay3676}
  {\bibfield  {journal} {\bibinfo  {journal} {Science}\ }\textbf {\bibinfo
  {volume} {369}},\ \bibinfo {pages} {eaay3676} (\bibinfo {year}
  {2020})}\BibitemShut {NoStop}%
\bibitem [{\citenamefont {Chang}\ \emph {et~al.}(2022)\citenamefont {Chang},
  \citenamefont {Liu},\ and\ \citenamefont {Bowers}}]{2022NP}%
  \BibitemOpen
  \bibfield  {author} {\bibinfo {author} {\bibfnamefont {L.}~\bibnamefont
  {Chang}}, \bibinfo {author} {\bibfnamefont {S.}~\bibnamefont {Liu}},\ and\
  \bibinfo {author} {\bibfnamefont {J.~E.}\ \bibnamefont {Bowers}},\ }\bibfield
   {title} {\bibinfo {title} {Integrated optical frequency comb technologies},\
  }\href {https://doi.org/10.1038/s41566-021-00945-1} {\bibfield  {journal}
  {\bibinfo  {journal} {Nat. Photonics}\ }\textbf {\bibinfo {volume} {16}},\
  \bibinfo {pages} {95} (\bibinfo {year} {2022})}\BibitemShut {NoStop}%
\bibitem [{\citenamefont {Little}\ \emph {et~al.}(1997)\citenamefont {Little},
  \citenamefont {Chu}, \citenamefont {Haus}, \citenamefont {Foresi},\ and\
  \citenamefont {Laine}}]{1997JLT}%
  \BibitemOpen
  \bibfield  {author} {\bibinfo {author} {\bibfnamefont {B.}~\bibnamefont
  {Little}}, \bibinfo {author} {\bibfnamefont {S.}~\bibnamefont {Chu}},
  \bibinfo {author} {\bibfnamefont {H.}~\bibnamefont {Haus}}, \bibinfo {author}
  {\bibfnamefont {J.}~\bibnamefont {Foresi}},\ and\ \bibinfo {author}
  {\bibfnamefont {J.-P.}\ \bibnamefont {Laine}},\ }\bibfield  {title} {\bibinfo
  {title} {Microring resonator channel dropping filters},\ }\href
  {https://doi.org/10.1109/50.588673} {\bibfield  {journal} {\bibinfo
  {journal} {J. Light. Technol.}\ }\textbf {\bibinfo {volume} {15}},\ \bibinfo
  {pages} {998} (\bibinfo {year} {1997})}\BibitemShut {NoStop}%
\bibitem [{\citenamefont {Klein}\ \emph {et~al.}(2005)\citenamefont {Klein},
  \citenamefont {Geuzebroek}, \citenamefont {Kelderman}, \citenamefont {Sengo},
  \citenamefont {Baker},\ and\ \citenamefont {Driessen}}]{2005IEEE}%
  \BibitemOpen
  \bibfield  {author} {\bibinfo {author} {\bibfnamefont {E.}~\bibnamefont
  {Klein}}, \bibinfo {author} {\bibfnamefont {D.}~\bibnamefont {Geuzebroek}},
  \bibinfo {author} {\bibfnamefont {H.}~\bibnamefont {Kelderman}}, \bibinfo
  {author} {\bibfnamefont {G.}~\bibnamefont {Sengo}}, \bibinfo {author}
  {\bibfnamefont {N.}~\bibnamefont {Baker}},\ and\ \bibinfo {author}
  {\bibfnamefont {A.}~\bibnamefont {Driessen}},\ }\bibfield  {title} {\bibinfo
  {title} {Reconfigurable optical add-drop multiplexer using microring
  resonators},\ }\href {https://doi.org/10.1109/LPT.2005.858131} {\bibfield
  {journal} {\bibinfo  {journal} {IEEE Photon. Technol. Lett.}\ }\textbf
  {\bibinfo {volume} {17}},\ \bibinfo {pages} {2358} (\bibinfo {year}
  {2005})}\BibitemShut {NoStop}%
\bibitem [{\citenamefont {Sherwood-Droz}\ \emph {et~al.}(2008)\citenamefont
  {Sherwood-Droz}, \citenamefont {Wang}, \citenamefont {Chen}, \citenamefont
  {Lee}, \citenamefont {Biberman}, \citenamefont {Bergman},\ and\ \citenamefont
  {Lipson}}]{2008OE2}%
  \BibitemOpen
  \bibfield  {author} {\bibinfo {author} {\bibfnamefont {N.}~\bibnamefont
  {Sherwood-Droz}}, \bibinfo {author} {\bibfnamefont {H.}~\bibnamefont {Wang}},
  \bibinfo {author} {\bibfnamefont {L.}~\bibnamefont {Chen}}, \bibinfo {author}
  {\bibfnamefont {B.~G.}\ \bibnamefont {Lee}}, \bibinfo {author} {\bibfnamefont
  {A.}~\bibnamefont {Biberman}}, \bibinfo {author} {\bibfnamefont
  {K.}~\bibnamefont {Bergman}},\ and\ \bibinfo {author} {\bibfnamefont
  {M.}~\bibnamefont {Lipson}},\ }\bibfield  {title} {\bibinfo {title} {Optical
  4{\texttimes}4 hitless silicon router for optical networks-on-chip
  ({N}o{C})},\ }\href {https://doi.org/10.1364/OE.16.015915} {\bibfield
  {journal} {\bibinfo  {journal} {Opt. Express}\ }\textbf {\bibinfo {volume}
  {16}},\ \bibinfo {pages} {15915} (\bibinfo {year} {2008})}\BibitemShut
  {NoStop}%
\bibitem [{\citenamefont {Wiederhecker}\ \emph {et~al.}(2011)\citenamefont
  {Wiederhecker}, \citenamefont {Manipatruni}, \citenamefont {Lee},\ and\
  \citenamefont {Lipson}}]{2011OE}%
  \BibitemOpen
  \bibfield  {author} {\bibinfo {author} {\bibfnamefont {G.~S.}\ \bibnamefont
  {Wiederhecker}}, \bibinfo {author} {\bibfnamefont {S.}~\bibnamefont
  {Manipatruni}}, \bibinfo {author} {\bibfnamefont {S.}~\bibnamefont {Lee}},\
  and\ \bibinfo {author} {\bibfnamefont {M.}~\bibnamefont {Lipson}},\
  }\bibfield  {title} {\bibinfo {title} {Broadband tuning of optomechanical
  cavities},\ }\href {https://doi.org/10.1364/OE.19.002782} {\bibfield
  {journal} {\bibinfo  {journal} {Opt. Express}\ }\textbf {\bibinfo {volume}
  {19}},\ \bibinfo {pages} {2782} (\bibinfo {year} {2011})}\BibitemShut
  {NoStop}%
\bibitem [{\citenamefont {Bekker}\ \emph {et~al.}(2018)\citenamefont {Bekker},
  \citenamefont {Baker}, \citenamefont {Kalra}, \citenamefont {Cheng},
  \citenamefont {Li}, \citenamefont {Prakash},\ and\ \citenamefont
  {Bowen}}]{2018OE}%
  \BibitemOpen
  \bibfield  {author} {\bibinfo {author} {\bibfnamefont {C.}~\bibnamefont
  {Bekker}}, \bibinfo {author} {\bibfnamefont {C.~G.}\ \bibnamefont {Baker}},
  \bibinfo {author} {\bibfnamefont {R.}~\bibnamefont {Kalra}}, \bibinfo
  {author} {\bibfnamefont {H.-H.}\ \bibnamefont {Cheng}}, \bibinfo {author}
  {\bibfnamefont {B.-B.}\ \bibnamefont {Li}}, \bibinfo {author} {\bibfnamefont
  {V.}~\bibnamefont {Prakash}},\ and\ \bibinfo {author} {\bibfnamefont {W.~P.}\
  \bibnamefont {Bowen}},\ }\bibfield  {title} {\bibinfo {title} {Free spectral
  range electrical tuning of a high quality on-chip microcavity},\ }\href
  {https://doi.org/10.1364/OE.26.033649} {\bibfield  {journal} {\bibinfo
  {journal} {Opt. Express}\ }\textbf {\bibinfo {volume} {26}},\ \bibinfo
  {pages} {33649} (\bibinfo {year} {2018})}\BibitemShut {NoStop}%
\bibitem [{\citenamefont {P\"ollinger}\ \emph {et~al.}(2009)\citenamefont
  {P\"ollinger}, \citenamefont {O'Shea}, \citenamefont {Warken},\ and\
  \citenamefont {Rauschenbeutel}}]{2009PRL}%
  \BibitemOpen
  \bibfield  {author} {\bibinfo {author} {\bibfnamefont {M.}~\bibnamefont
  {P\"ollinger}}, \bibinfo {author} {\bibfnamefont {D.}~\bibnamefont {O'Shea}},
  \bibinfo {author} {\bibfnamefont {F.}~\bibnamefont {Warken}},\ and\ \bibinfo
  {author} {\bibfnamefont {A.}~\bibnamefont {Rauschenbeutel}},\ }\bibfield
  {title} {\bibinfo {title} {Ultrahigh-{$Q$} tunable whispering-gallery-mode
  microresonator},\ }\href {https://doi.org/10.1103/PhysRevLett.103.053901}
  {\bibfield  {journal} {\bibinfo  {journal} {Phys. Rev. Lett.}\ }\textbf
  {\bibinfo {volume} {103}},\ \bibinfo {pages} {053901} (\bibinfo {year}
  {2009})}\BibitemShut {NoStop}%
\bibitem [{\citenamefont {Sumetsky}\ \emph {et~al.}(2010)\citenamefont
  {Sumetsky}, \citenamefont {Dulashko},\ and\ \citenamefont
  {Windeler}}]{2010OL}%
  \BibitemOpen
  \bibfield  {author} {\bibinfo {author} {\bibfnamefont {M.}~\bibnamefont
  {Sumetsky}}, \bibinfo {author} {\bibfnamefont {Y.}~\bibnamefont {Dulashko}},\
  and\ \bibinfo {author} {\bibfnamefont {R.~S.}\ \bibnamefont {Windeler}},\
  }\bibfield  {title} {\bibinfo {title} {Super free spectral range tunable
  optical microbubble resonator},\ }\href
  {https://doi.org/10.1364/OL.35.001866} {\bibfield  {journal} {\bibinfo
  {journal} {Opt. Lett.}\ }\textbf {\bibinfo {volume} {35}},\ \bibinfo {pages}
  {1866} (\bibinfo {year} {2010})}\BibitemShut {NoStop}%
\bibitem [{\citenamefont {Jin}\ \emph {et~al.}(2018)\citenamefont {Jin},
  \citenamefont {Polcawich}, \citenamefont {Morton},\ and\ \citenamefont
  {Bowers}}]{2018OE2}%
  \BibitemOpen
  \bibfield  {author} {\bibinfo {author} {\bibfnamefont {W.}~\bibnamefont
  {Jin}}, \bibinfo {author} {\bibfnamefont {R.~G.}\ \bibnamefont {Polcawich}},
  \bibinfo {author} {\bibfnamefont {P.~A.}\ \bibnamefont {Morton}},\ and\
  \bibinfo {author} {\bibfnamefont {J.~E.}\ \bibnamefont {Bowers}},\ }\bibfield
   {title} {\bibinfo {title} {Piezoelectrically tuned silicon nitride ring
  resonator},\ }\href {https://doi.org/10.1364/OE.26.003174} {\bibfield
  {journal} {\bibinfo  {journal} {Opt. Express}\ }\textbf {\bibinfo {volume}
  {26}},\ \bibinfo {pages} {3174} (\bibinfo {year} {2018})}\BibitemShut
  {NoStop}%
\bibitem [{\citenamefont {Ikeda}\ and\ \citenamefont {Hane}(2013)}]{2013APL2}%
  \BibitemOpen
  \bibfield  {author} {\bibinfo {author} {\bibfnamefont {T.}~\bibnamefont
  {Ikeda}}\ and\ \bibinfo {author} {\bibfnamefont {K.}~\bibnamefont {Hane}},\
  }\bibfield  {title} {\bibinfo {title} {A microelectromechanically tunable
  microring resonator composed of freestanding silicon photonic waveguide
  couplers},\ }\href {https://doi.org/10.1063/1.4809733} {\bibfield  {journal}
  {\bibinfo  {journal} {Appl. Phys. Lett.}\ }\textbf {\bibinfo {volume}
  {102}},\ \bibinfo {pages} {221113} (\bibinfo {year} {2013})}\BibitemShut
  {NoStop}%
\bibitem [{\citenamefont {Zhu}\ \emph {et~al.}(2017)\citenamefont {Zhu},
  \citenamefont {Shi}, \citenamefont {Yuan}, \citenamefont {Xu},\ and\
  \citenamefont {Zhang}}]{2017OL}%
  \BibitemOpen
  \bibfield  {author} {\bibinfo {author} {\bibfnamefont {S.}~\bibnamefont
  {Zhu}}, \bibinfo {author} {\bibfnamefont {L.}~\bibnamefont {Shi}}, \bibinfo
  {author} {\bibfnamefont {S.}~\bibnamefont {Yuan}}, \bibinfo {author}
  {\bibfnamefont {X.}~\bibnamefont {Xu}},\ and\ \bibinfo {author}
  {\bibfnamefont {X.}~\bibnamefont {Zhang}},\ }\bibfield  {title} {\bibinfo
  {title} {All-optical control of ultrahigh-{Q} silica microcavities with iron
  oxide nanoparticles},\ }\href {https://doi.org/10.1364/OL.42.005133}
  {\bibfield  {journal} {\bibinfo  {journal} {Opt. Lett.}\ }\textbf {\bibinfo
  {volume} {42}},\ \bibinfo {pages} {5133} (\bibinfo {year}
  {2017})}\BibitemShut {NoStop}%
\bibitem [{\citenamefont {Tang}\ \emph {et~al.}(2018)\citenamefont {Tang},
  \citenamefont {Liu}, \citenamefont {Qian}, \citenamefont {Shi}, \citenamefont
  {Sun}, \citenamefont {Wu}, \citenamefont {Gong},\ and\ \citenamefont
  {Xiao}}]{2018AM}%
  \BibitemOpen
  \bibfield  {author} {\bibinfo {author} {\bibfnamefont {S.-J.}\ \bibnamefont
  {Tang}}, \bibinfo {author} {\bibfnamefont {Z.}~\bibnamefont {Liu}}, \bibinfo
  {author} {\bibfnamefont {Y.-J.}\ \bibnamefont {Qian}}, \bibinfo {author}
  {\bibfnamefont {K.}~\bibnamefont {Shi}}, \bibinfo {author} {\bibfnamefont
  {Y.}~\bibnamefont {Sun}}, \bibinfo {author} {\bibfnamefont {C.}~\bibnamefont
  {Wu}}, \bibinfo {author} {\bibfnamefont {Q.}~\bibnamefont {Gong}},\ and\
  \bibinfo {author} {\bibfnamefont {Y.-F.}\ \bibnamefont {Xiao}},\ }\bibfield
  {title} {\bibinfo {title} {A tunable optofluidic microlaser in a photostable
  conjugated polymer},\ }\href
  {https://doi.org/https://doi.org/10.1002/adma.201804556} {\bibfield
  {journal} {\bibinfo  {journal} {Adv. Mater.}\ }\textbf {\bibinfo {volume}
  {30}},\ \bibinfo {pages} {1804556} (\bibinfo {year} {2018})}\BibitemShut
  {NoStop}%
\bibitem [{\citenamefont {Zhu}\ \emph {et~al.}(2018)\citenamefont {Zhu},
  \citenamefont {Shi}, \citenamefont {Xiao}, \citenamefont {Zhang},\ and\
  \citenamefont {Fan}}]{2018ACS}%
  \BibitemOpen
  \bibfield  {author} {\bibinfo {author} {\bibfnamefont {S.}~\bibnamefont
  {Zhu}}, \bibinfo {author} {\bibfnamefont {L.}~\bibnamefont {Shi}}, \bibinfo
  {author} {\bibfnamefont {B.}~\bibnamefont {Xiao}}, \bibinfo {author}
  {\bibfnamefont {X.}~\bibnamefont {Zhang}},\ and\ \bibinfo {author}
  {\bibfnamefont {X.}~\bibnamefont {Fan}},\ }\bibfield  {title} {\bibinfo
  {title} {All-optical tunable microlaser based on an ultrahigh-{Q}
  erbium-doped hybrid microbottle cavity},\ }\href
  {https://doi.org/10.1021/acsphotonics.8b00838} {\bibfield  {journal}
  {\bibinfo  {journal} {ACS Photonics}\ }\textbf {\bibinfo {volume} {5}},\
  \bibinfo {pages} {3794} (\bibinfo {year} {2018})}\BibitemShut {NoStop}%
\bibitem [{\citenamefont {Perahia}\ \emph {et~al.}(2010)\citenamefont
  {Perahia}, \citenamefont {Cohen}, \citenamefont {Meenehan}, \citenamefont
  {Alegre},\ and\ \citenamefont {Painter}}]{2010APL}%
  \BibitemOpen
  \bibfield  {author} {\bibinfo {author} {\bibfnamefont {R.}~\bibnamefont
  {Perahia}}, \bibinfo {author} {\bibfnamefont {J.~D.}\ \bibnamefont {Cohen}},
  \bibinfo {author} {\bibfnamefont {S.}~\bibnamefont {Meenehan}}, \bibinfo
  {author} {\bibfnamefont {T.~P.~M.}\ \bibnamefont {Alegre}},\ and\ \bibinfo
  {author} {\bibfnamefont {O.}~\bibnamefont {Painter}},\ }\bibfield  {title}
  {\bibinfo {title} {Electrostatically tunable optomechanical "zipper"
  cavity laser},\ }\href {https://doi.org/10.1063/1.3515296} {\bibfield
  {journal} {\bibinfo  {journal} {Appl. Phys. Lett.}\ }\textbf {\bibinfo
  {volume} {97}},\ \bibinfo {pages} {191112} (\bibinfo {year}
  {2010})}\BibitemShut {NoStop}%
\bibitem [{\citenamefont {Yang}\ \emph {et~al.}(2017)\citenamefont {Yang},
  \citenamefont {Lei}, \citenamefont {Kasumie}, \citenamefont {Xu},
  \citenamefont {Ward}, \citenamefont {Yang},\ and\ \citenamefont
  {Chormaic}}]{2017OE}%
  \BibitemOpen
  \bibfield  {author} {\bibinfo {author} {\bibfnamefont {Y.}~\bibnamefont
  {Yang}}, \bibinfo {author} {\bibfnamefont {F.}~\bibnamefont {Lei}}, \bibinfo
  {author} {\bibfnamefont {S.}~\bibnamefont {Kasumie}}, \bibinfo {author}
  {\bibfnamefont {L.}~\bibnamefont {Xu}}, \bibinfo {author} {\bibfnamefont
  {J.~M.}\ \bibnamefont {Ward}}, \bibinfo {author} {\bibfnamefont
  {L.}~\bibnamefont {Yang}},\ and\ \bibinfo {author} {\bibfnamefont {S.~N.}\
  \bibnamefont {Chormaic}},\ }\bibfield  {title} {\bibinfo {title} {Tunable
  erbium-doped microbubble laser fabricated by sol-gel coating},\ }\href
  {https://doi.org/10.1364/OE.25.001308} {\bibfield  {journal} {\bibinfo
  {journal} {Opt. Express}\ }\textbf {\bibinfo {volume} {25}},\ \bibinfo
  {pages} {1308} (\bibinfo {year} {2017})}\BibitemShut {NoStop}%
\bibitem [{\citenamefont {Ward}\ \emph {et~al.}(2016)\citenamefont {Ward},
  \citenamefont {Yang},\ and\ \citenamefont {N.~Chormaic}}]{2016SR}%
  \BibitemOpen
  \bibfield  {author} {\bibinfo {author} {\bibfnamefont {J.~M.}\ \bibnamefont
  {Ward}}, \bibinfo {author} {\bibfnamefont {Y.}~\bibnamefont {Yang}},\ and\
  \bibinfo {author} {\bibfnamefont {S.}~\bibnamefont {N.~Chormaic}},\
  }\bibfield  {title} {\bibinfo {title} {Glass-on-glass fabrication of
  bottle-shaped tunable microlasers and their applications},\ }\href
  {https://doi.org/10.1038/srep25152} {\bibfield  {journal} {\bibinfo
  {journal} {Sci. Rep.}\ }\textbf {\bibinfo {volume} {6}},\ \bibinfo {pages}
  {25152} (\bibinfo {year} {2016})}\BibitemShut {NoStop}%
\bibitem [{\citenamefont {Ma}\ \emph {et~al.}(2018)\citenamefont {Ma},
  \citenamefont {Yuan}, \citenamefont {Zhu}, \citenamefont {Shi},\ and\
  \citenamefont {Zhang}}]{2018OL}%
  \BibitemOpen
  \bibfield  {author} {\bibinfo {author} {\bibfnamefont {R.}~\bibnamefont
  {Ma}}, \bibinfo {author} {\bibfnamefont {S.}~\bibnamefont {Yuan}}, \bibinfo
  {author} {\bibfnamefont {S.}~\bibnamefont {Zhu}}, \bibinfo {author}
  {\bibfnamefont {L.}~\bibnamefont {Shi}},\ and\ \bibinfo {author}
  {\bibfnamefont {X.}~\bibnamefont {Zhang}},\ }\bibfield  {title} {\bibinfo
  {title} {Tunable sub-k{H}z single-mode fiber laser based on a hybrid
  microbottle resonator},\ }\href {https://doi.org/10.1364/OL.43.005315}
  {\bibfield  {journal} {\bibinfo  {journal} {Opt. Lett.}\ }\textbf {\bibinfo
  {volume} {43}},\ \bibinfo {pages} {5315} (\bibinfo {year}
  {2018})}\BibitemShut {NoStop}%
\bibitem [{\citenamefont {Zhu}\ \emph {et~al.}(2019)\citenamefont {Zhu},
  \citenamefont {Xiao}, \citenamefont {Jiang}, \citenamefont {Shi},\ and\
  \citenamefont {Zhang}}]{2019nanoph}%
  \BibitemOpen
  \bibfield  {author} {\bibinfo {author} {\bibfnamefont {S.}~\bibnamefont
  {Zhu}}, \bibinfo {author} {\bibfnamefont {B.}~\bibnamefont {Xiao}}, \bibinfo
  {author} {\bibfnamefont {B.}~\bibnamefont {Jiang}}, \bibinfo {author}
  {\bibfnamefont {L.}~\bibnamefont {Shi}},\ and\ \bibinfo {author}
  {\bibfnamefont {X.}~\bibnamefont {Zhang}},\ }\bibfield  {title} {\bibinfo
  {title} {Tunable {B}rillouin and {R}aman microlasers using hybrid microbottle
  resonators},\ }\href {https://doi.org/doi:10.1515/nanoph-2019-0070}
  {\bibfield  {journal} {\bibinfo  {journal} {Nanophotonics}\ }\textbf
  {\bibinfo {volume} {8}},\ \bibinfo {pages} {931} (\bibinfo {year}
  {2019})}\BibitemShut {NoStop}%
\bibitem [{\citenamefont {Armani}\ \emph {et~al.}(2004)\citenamefont {Armani},
  \citenamefont {Min}, \citenamefont {Martin},\ and\ \citenamefont
  {Vahala}}]{2004APL}%
  \BibitemOpen
  \bibfield  {author} {\bibinfo {author} {\bibfnamefont {D.}~\bibnamefont
  {Armani}}, \bibinfo {author} {\bibfnamefont {B.}~\bibnamefont {Min}},
  \bibinfo {author} {\bibfnamefont {A.}~\bibnamefont {Martin}},\ and\ \bibinfo
  {author} {\bibfnamefont {K.~J.}\ \bibnamefont {Vahala}},\ }\bibfield  {title}
  {\bibinfo {title} {Electrical thermo-optic tuning of ultrahigh-{Q}
  microtoroid resonators},\ }\href {https://doi.org/10.1063/1.1825069}
  {\bibfield  {journal} {\bibinfo  {journal} {Appl. Phys. Lett.}\ }\textbf
  {\bibinfo {volume} {85}},\ \bibinfo {pages} {5439} (\bibinfo {year}
  {2004})}\BibitemShut {NoStop}%
\bibitem [{\citenamefont {Zhou}\ \emph {et~al.}(2013)\citenamefont {Zhou},
  \citenamefont {Zhang}, \citenamefont {Lu},\ and\ \citenamefont
  {Chen}}]{2013IEEE}%
  \BibitemOpen
  \bibfield  {author} {\bibinfo {author} {\bibfnamefont {L.}~\bibnamefont
  {Zhou}}, \bibinfo {author} {\bibfnamefont {X.}~\bibnamefont {Zhang}},
  \bibinfo {author} {\bibfnamefont {L.}~\bibnamefont {Lu}},\ and\ \bibinfo
  {author} {\bibfnamefont {J.}~\bibnamefont {Chen}},\ }\bibfield  {title}
  {\bibinfo {title} {Tunable {V}ernier microring optical filters with
  $p-i-p$-type microheaters},\ }\href
  {https://doi.org/10.1109/JPHOT.2013.2271901} {\bibfield  {journal} {\bibinfo
  {journal} {IEEE Photon. J.}\ }\textbf {\bibinfo {volume} {5}},\ \bibinfo
  {pages} {6601211} (\bibinfo {year} {2013})}\BibitemShut {NoStop}%
\bibitem [{\citenamefont {Lee}\ \emph {et~al.}(2017)\citenamefont {Lee},
  \citenamefont {Zhang}, \citenamefont {Barbosa}, \citenamefont {Miller},
  \citenamefont {Mohanty}, \citenamefont {St-Gelais},\ and\ \citenamefont
  {Lipson}}]{2017OE3}%
  \BibitemOpen
  \bibfield  {author} {\bibinfo {author} {\bibfnamefont {B.~S.}\ \bibnamefont
  {Lee}}, \bibinfo {author} {\bibfnamefont {M.}~\bibnamefont {Zhang}}, \bibinfo
  {author} {\bibfnamefont {F.~A.~S.}\ \bibnamefont {Barbosa}}, \bibinfo
  {author} {\bibfnamefont {S.~A.}\ \bibnamefont {Miller}}, \bibinfo {author}
  {\bibfnamefont {A.}~\bibnamefont {Mohanty}}, \bibinfo {author} {\bibfnamefont
  {R.}~\bibnamefont {St-Gelais}},\ and\ \bibinfo {author} {\bibfnamefont
  {M.}~\bibnamefont {Lipson}},\ }\bibfield  {title} {\bibinfo {title} {On-chip
  thermo-optic tuning of suspended microresonators},\ }\href
  {https://doi.org/10.1364/OE.25.012109} {\bibfield  {journal} {\bibinfo
  {journal} {Opt. Express}\ }\textbf {\bibinfo {volume} {25}},\ \bibinfo
  {pages} {12109} (\bibinfo {year} {2017})}\BibitemShut {NoStop}%
\bibitem [{\citenamefont {Wang}\ \emph {et~al.}(2007)\citenamefont {Wang},
  \citenamefont {Chu},\ and\ \citenamefont {Lin}}]{2007OL}%
  \BibitemOpen
  \bibfield  {author} {\bibinfo {author} {\bibfnamefont {T.-J.}\ \bibnamefont
  {Wang}}, \bibinfo {author} {\bibfnamefont {C.-H.}\ \bibnamefont {Chu}},\ and\
  \bibinfo {author} {\bibfnamefont {C.-Y.}\ \bibnamefont {Lin}},\ }\bibfield
  {title} {\bibinfo {title} {Electro-optically tunable microring resonators on
  lithium niobate},\ }\href {https://doi.org/10.1364/OL.32.002777} {\bibfield
  {journal} {\bibinfo  {journal} {Opt. Lett.}\ }\textbf {\bibinfo {volume}
  {32}},\ \bibinfo {pages} {2777} (\bibinfo {year} {2007})}\BibitemShut
  {NoStop}%
\bibitem [{\citenamefont {Guarino}\ \emph {et~al.}(2007)\citenamefont
  {Guarino}, \citenamefont {Poberaj}, \citenamefont {Rezzonico}, \citenamefont
  {Degl'Innocenti},\ and\ \citenamefont {Günter}}]{2007NP2}%
  \BibitemOpen
  \bibfield  {author} {\bibinfo {author} {\bibfnamefont {A.}~\bibnamefont
  {Guarino}}, \bibinfo {author} {\bibfnamefont {G.}~\bibnamefont {Poberaj}},
  \bibinfo {author} {\bibfnamefont {D.}~\bibnamefont {Rezzonico}}, \bibinfo
  {author} {\bibfnamefont {R.}~\bibnamefont {Degl'Innocenti}},\ and\ \bibinfo
  {author} {\bibfnamefont {P.}~\bibnamefont {Günter}},\ }\bibfield  {title}
  {\bibinfo {title} {Electro–optically tunable microring resonators in
  lithium niobate},\ }\href {https://doi.org/10.1038/nphoton.2007.93}
  {\bibfield  {journal} {\bibinfo  {journal} {Nat. Photonics}\ }\textbf
  {\bibinfo {volume} {1}},\ \bibinfo {pages} {407} (\bibinfo {year}
  {2007})}\BibitemShut {NoStop}%
\bibitem [{\citenamefont {Djordjev}\ \emph {et~al.}(2002)\citenamefont
  {Djordjev}, \citenamefont {Choi}, \citenamefont {Choi},\ and\ \citenamefont
  {Dapkus}}]{2002IEEE2}%
  \BibitemOpen
  \bibfield  {author} {\bibinfo {author} {\bibfnamefont {K.}~\bibnamefont
  {Djordjev}}, \bibinfo {author} {\bibfnamefont {S.-J.}\ \bibnamefont {Choi}},
  \bibinfo {author} {\bibfnamefont {S.-J.}\ \bibnamefont {Choi}},\ and\
  \bibinfo {author} {\bibfnamefont {R.}~\bibnamefont {Dapkus}},\ }\bibfield
  {title} {\bibinfo {title} {Microdisk tunable resonant filters and switches},\
  }\href {https://doi.org/10.1109/LPT.2002.1003107} {\bibfield  {journal}
  {\bibinfo  {journal} {IEEE Photon. Technol. Lett.}\ }\textbf {\bibinfo
  {volume} {14}},\ \bibinfo {pages} {828} (\bibinfo {year} {2002})}\BibitemShut
  {NoStop}%
\bibitem [{\citenamefont {Lipson}(2006)}]{2006IEEE}%
  \BibitemOpen
  \bibfield  {author} {\bibinfo {author} {\bibfnamefont {M.}~\bibnamefont
  {Lipson}},\ }\bibfield  {title} {\bibinfo {title} {Compact electro-optic
  modulators on a silicon chip},\ }\href
  {https://doi.org/10.1109/JSTQE.2006.885341} {\bibfield  {journal} {\bibinfo
  {journal} {IEEE J. Sel. Top. Quantum Electron.}\ }\textbf {\bibinfo {volume}
  {12}},\ \bibinfo {pages} {1520} (\bibinfo {year} {2006})}\BibitemShut
  {NoStop}%
\bibitem [{\citenamefont {Henze}\ \emph {et~al.}(2011)\citenamefont {Henze},
  \citenamefont {Seifert}, \citenamefont {Ward},\ and\ \citenamefont
  {Benson}}]{2011OL}%
  \BibitemOpen
  \bibfield  {author} {\bibinfo {author} {\bibfnamefont {R.}~\bibnamefont
  {Henze}}, \bibinfo {author} {\bibfnamefont {T.}~\bibnamefont {Seifert}},
  \bibinfo {author} {\bibfnamefont {J.}~\bibnamefont {Ward}},\ and\ \bibinfo
  {author} {\bibfnamefont {O.}~\bibnamefont {Benson}},\ }\bibfield  {title}
  {\bibinfo {title} {Tuning whispering gallery modes using internal aerostatic
  pressure},\ }\href {https://doi.org/10.1364/OL.36.004536} {\bibfield
  {journal} {\bibinfo  {journal} {Opt. Lett.}\ }\textbf {\bibinfo {volume}
  {36}},\ \bibinfo {pages} {4536} (\bibinfo {year} {2011})}\BibitemShut
  {NoStop}%
\bibitem [{\citenamefont {Dinyari}\ \emph {et~al.}(2011)\citenamefont
  {Dinyari}, \citenamefont {Barbour}, \citenamefont {Golter},\ and\
  \citenamefont {Wang}}]{2011OE2}%
  \BibitemOpen
  \bibfield  {author} {\bibinfo {author} {\bibfnamefont {K.~N.}\ \bibnamefont
  {Dinyari}}, \bibinfo {author} {\bibfnamefont {R.~J.}\ \bibnamefont
  {Barbour}}, \bibinfo {author} {\bibfnamefont {D.~A.}\ \bibnamefont
  {Golter}},\ and\ \bibinfo {author} {\bibfnamefont {H.}~\bibnamefont {Wang}},\
  }\bibfield  {title} {\bibinfo {title} {Mechanical tuning of whispering
  gallery modes over a 0.5 {TH}z tuning range with {MH}z resolution in a silica
  microsphere at cryogenic temperatures},\ }\href
  {https://doi.org/10.1364/OE.19.017966} {\bibfield  {journal} {\bibinfo
  {journal} {Opt. Express}\ }\textbf {\bibinfo {volume} {19}},\ \bibinfo
  {pages} {17966} (\bibinfo {year} {2011})}\BibitemShut {NoStop}%
\bibitem [{\citenamefont {Baker}\ \emph {et~al.}(2016)\citenamefont {Baker},
  \citenamefont {Bekker}, \citenamefont {McAuslan}, \citenamefont {Sheridan},\
  and\ \citenamefont {Bowen}}]{2016OE}%
  \BibitemOpen
  \bibfield  {author} {\bibinfo {author} {\bibfnamefont {C.~G.}\ \bibnamefont
  {Baker}}, \bibinfo {author} {\bibfnamefont {C.}~\bibnamefont {Bekker}},
  \bibinfo {author} {\bibfnamefont {D.~L.}\ \bibnamefont {McAuslan}}, \bibinfo
  {author} {\bibfnamefont {E.}~\bibnamefont {Sheridan}},\ and\ \bibinfo
  {author} {\bibfnamefont {W.~P.}\ \bibnamefont {Bowen}},\ }\bibfield  {title}
  {\bibinfo {title} {High bandwidth on-chip capacitive tuning of microtoroid
  resonators},\ }\href {https://doi.org/10.1364/OE.24.020400} {\bibfield
  {journal} {\bibinfo  {journal} {Opt. Express}\ }\textbf {\bibinfo {volume}
  {24}},\ \bibinfo {pages} {20400} (\bibinfo {year} {2016})}\BibitemShut
  {NoStop}%
\bibitem [{\citenamefont {Chen}\ \emph {et~al.}(2017)\citenamefont {Chen},
  \citenamefont {Zhou}, \citenamefont {Zou}, \citenamefont {Shen},
  \citenamefont {Guo},\ and\ \citenamefont {Dong}}]{2017OE2}%
  \BibitemOpen
  \bibfield  {author} {\bibinfo {author} {\bibfnamefont {Y.}~\bibnamefont
  {Chen}}, \bibinfo {author} {\bibfnamefont {Z.-H.}\ \bibnamefont {Zhou}},
  \bibinfo {author} {\bibfnamefont {C.-L.}\ \bibnamefont {Zou}}, \bibinfo
  {author} {\bibfnamefont {Z.}~\bibnamefont {Shen}}, \bibinfo {author}
  {\bibfnamefont {G.-C.}\ \bibnamefont {Guo}},\ and\ \bibinfo {author}
  {\bibfnamefont {C.-H.}\ \bibnamefont {Dong}},\ }\bibfield  {title} {\bibinfo
  {title} {Tunable {R}aman laser in a hollow bottle-like microresonator},\
  }\href {https://doi.org/10.1364/OE.25.016879} {\bibfield  {journal} {\bibinfo
   {journal} {Opt. Express}\ }\textbf {\bibinfo {volume} {25}},\ \bibinfo
  {pages} {16879} (\bibinfo {year} {2017})}\BibitemShut {NoStop}%
\bibitem [{\citenamefont {Liu}\ \emph {et~al.}(2020)\citenamefont {Liu},
  \citenamefont {Tian}, \citenamefont {Lucas}, \citenamefont {Raja},
  \citenamefont {Lihachev}, \citenamefont {Wang}, \citenamefont {He},
  \citenamefont {Liu}, \citenamefont {Anderson}, \citenamefont {Weng},
  \citenamefont {Bhave},\ and\ \citenamefont {Kippenberg}}]{2020N}%
  \BibitemOpen
  \bibfield  {author} {\bibinfo {author} {\bibfnamefont {J.}~\bibnamefont
  {Liu}}, \bibinfo {author} {\bibfnamefont {H.}~\bibnamefont {Tian}}, \bibinfo
  {author} {\bibfnamefont {E.}~\bibnamefont {Lucas}}, \bibinfo {author}
  {\bibfnamefont {A.~S.}\ \bibnamefont {Raja}}, \bibinfo {author}
  {\bibfnamefont {G.}~\bibnamefont {Lihachev}}, \bibinfo {author}
  {\bibfnamefont {R.~N.}\ \bibnamefont {Wang}}, \bibinfo {author}
  {\bibfnamefont {J.}~\bibnamefont {He}}, \bibinfo {author} {\bibfnamefont
  {T.}~\bibnamefont {Liu}}, \bibinfo {author} {\bibfnamefont {M.~H.}\
  \bibnamefont {Anderson}}, \bibinfo {author} {\bibfnamefont {W.}~\bibnamefont
  {Weng}}, \bibinfo {author} {\bibfnamefont {S.~A.}\ \bibnamefont {Bhave}},\
  and\ \bibinfo {author} {\bibfnamefont {T.~J.}\ \bibnamefont {Kippenberg}},\
  }\bibfield  {title} {\bibinfo {title} {Monolithic piezoelectric control of
  soliton microcombs},\ }\href {https://doi.org/10.1038/s41586-020-2465-8}
  {\bibfield  {journal} {\bibinfo  {journal} {Nature}\ }\textbf {\bibinfo
  {volume} {583}},\ \bibinfo {pages} {385} (\bibinfo {year}
  {2020})}\BibitemShut {NoStop}%
\bibitem [{\citenamefont {Rakich}\ \emph {et~al.}(2007)\citenamefont {Rakich},
  \citenamefont {Popović}, \citenamefont {Soljačić},\ and\ \citenamefont
  {Ippen}}]{2007NP}%
  \BibitemOpen
  \bibfield  {author} {\bibinfo {author} {\bibfnamefont {P.~T.}\ \bibnamefont
  {Rakich}}, \bibinfo {author} {\bibfnamefont {M.~A.}\ \bibnamefont
  {Popović}}, \bibinfo {author} {\bibfnamefont {M.}~\bibnamefont
  {Soljačić}},\ and\ \bibinfo {author} {\bibfnamefont {E.~P.}\ \bibnamefont
  {Ippen}},\ }\bibfield  {title} {\bibinfo {title} {Trapping, corralling and
  spectral bonding of optical resonances through optically induced
  potentials},\ }\href {https://doi.org/10.1038/nphoton.2007.203} {\bibfield
  {journal} {\bibinfo  {journal} {Nat. Photonics}\ }\textbf {\bibinfo {volume}
  {1}},\ \bibinfo {pages} {658} (\bibinfo {year} {2007})}\BibitemShut {NoStop}%
\bibitem [{\citenamefont {Rosenberg}\ \emph {et~al.}(2009)\citenamefont
  {Rosenberg}, \citenamefont {Lin},\ and\ \citenamefont {Painter}}]{2009NP}%
  \BibitemOpen
  \bibfield  {author} {\bibinfo {author} {\bibfnamefont {J.}~\bibnamefont
  {Rosenberg}}, \bibinfo {author} {\bibfnamefont {Q.}~\bibnamefont {Lin}},\
  and\ \bibinfo {author} {\bibfnamefont {O.}~\bibnamefont {Painter}},\
  }\bibfield  {title} {\bibinfo {title} {Static and dynamic wavelength routing
  via the gradient optical force},\ }\href
  {https://doi.org/10.1038/nphoton.2009.137} {\bibfield  {journal} {\bibinfo
  {journal} {Nat. Photonics}\ }\textbf {\bibinfo {volume} {3}},\ \bibinfo
  {pages} {478} (\bibinfo {year} {2009})}\BibitemShut {NoStop}%
\bibitem [{\citenamefont {Wiederhecker}\ \emph {et~al.}(2009)\citenamefont
  {Wiederhecker}, \citenamefont {Chen}, \citenamefont {Gondarenko},\ and\
  \citenamefont {Lipson}}]{2009N}%
  \BibitemOpen
  \bibfield  {author} {\bibinfo {author} {\bibfnamefont {G.~S.}\ \bibnamefont
  {Wiederhecker}}, \bibinfo {author} {\bibfnamefont {L.}~\bibnamefont {Chen}},
  \bibinfo {author} {\bibfnamefont {A.}~\bibnamefont {Gondarenko}},\ and\
  \bibinfo {author} {\bibfnamefont {M.}~\bibnamefont {Lipson}},\ }\bibfield
  {title} {\bibinfo {title} {Controlling photonic structures using optical
  forces},\ }\href {https://doi.org/10.1038/nature08584} {\bibfield  {journal}
  {\bibinfo  {journal} {Nature}\ }\textbf {\bibinfo {volume} {462}},\ \bibinfo
  {pages} {633} (\bibinfo {year} {2009})}\BibitemShut {NoStop}%
\bibitem [{\citenamefont {Jiang}\ \emph {et~al.}(2009)\citenamefont {Jiang},
  \citenamefont {Lin}, \citenamefont {Rosenberg}, \citenamefont {Vahala},\ and\
  \citenamefont {Painter}}]{2009OE}%
  \BibitemOpen
  \bibfield  {author} {\bibinfo {author} {\bibfnamefont {X.}~\bibnamefont
  {Jiang}}, \bibinfo {author} {\bibfnamefont {Q.}~\bibnamefont {Lin}}, \bibinfo
  {author} {\bibfnamefont {J.}~\bibnamefont {Rosenberg}}, \bibinfo {author}
  {\bibfnamefont {K.}~\bibnamefont {Vahala}},\ and\ \bibinfo {author}
  {\bibfnamefont {O.}~\bibnamefont {Painter}},\ }\bibfield  {title} {\bibinfo
  {title} {High-{Q} double-disk microcavities for cavity optomechanics},\
  }\href {https://doi.org/10.1364/OE.17.020911} {\bibfield  {journal} {\bibinfo
   {journal} {Opt. Express}\ }\textbf {\bibinfo {volume} {17}},\ \bibinfo
  {pages} {20911} (\bibinfo {year} {2009})}\BibitemShut {NoStop}%
\bibitem [{\citenamefont {Bekker}\ \emph {et~al.}(2021)\citenamefont {Bekker},
  \citenamefont {Baker},\ and\ \citenamefont {Bowen}}]{2021PRapp}%
  \BibitemOpen
  \bibfield  {author} {\bibinfo {author} {\bibfnamefont {C.~J.}\ \bibnamefont
  {Bekker}}, \bibinfo {author} {\bibfnamefont {C.~G.}\ \bibnamefont {Baker}},\
  and\ \bibinfo {author} {\bibfnamefont {W.~P.}\ \bibnamefont {Bowen}},\
  }\bibfield  {title} {\bibinfo {title} {Optically tunable photoluminescence
  and up-conversion lasing on a chip},\ }\href
  {https://doi.org/10.1103/PhysRevApplied.15.034022} {\bibfield  {journal}
  {\bibinfo  {journal} {Phys. Rev. Applied}\ }\textbf {\bibinfo {volume}
  {15}},\ \bibinfo {pages} {034022} (\bibinfo {year} {2021})}\BibitemShut
  {NoStop}%
\bibitem [{\citenamefont {Frank}\ \emph {et~al.}(2010)\citenamefont {Frank},
  \citenamefont {Deotare}, \citenamefont {McCutcheon},\ and\ \citenamefont
  {Lon\v{c}ar}}]{2010OE}%
  \BibitemOpen
  \bibfield  {author} {\bibinfo {author} {\bibfnamefont {I.~W.}\ \bibnamefont
  {Frank}}, \bibinfo {author} {\bibfnamefont {P.~B.}\ \bibnamefont {Deotare}},
  \bibinfo {author} {\bibfnamefont {M.~W.}\ \bibnamefont {McCutcheon}},\ and\
  \bibinfo {author} {\bibfnamefont {M.}~\bibnamefont {Lon\v{c}ar}},\ }\bibfield
   {title} {\bibinfo {title} {Programmable photonic crystal nanobeam
  cavities},\ }\href {https://doi.org/10.1364/OE.18.008705} {\bibfield
  {journal} {\bibinfo  {journal} {Opt. Express}\ }\textbf {\bibinfo {volume}
  {18}},\ \bibinfo {pages} {8705} (\bibinfo {year} {2010})}\BibitemShut
  {NoStop}%
\bibitem [{\citenamefont {Winger}\ \emph {et~al.}(2011)\citenamefont {Winger},
  \citenamefont {Blasius}, \citenamefont {Alegre}, \citenamefont
  {Safavi-Naeini}, \citenamefont {Meenehan}, \citenamefont {Cohen},
  \citenamefont {Stobbe},\ and\ \citenamefont {Painter}}]{2011OE3}%
  \BibitemOpen
  \bibfield  {author} {\bibinfo {author} {\bibfnamefont {M.}~\bibnamefont
  {Winger}}, \bibinfo {author} {\bibfnamefont {T.~D.}\ \bibnamefont {Blasius}},
  \bibinfo {author} {\bibfnamefont {T.~P.~M.}\ \bibnamefont {Alegre}}, \bibinfo
  {author} {\bibfnamefont {A.~H.}\ \bibnamefont {Safavi-Naeini}}, \bibinfo
  {author} {\bibfnamefont {S.}~\bibnamefont {Meenehan}}, \bibinfo {author}
  {\bibfnamefont {J.}~\bibnamefont {Cohen}}, \bibinfo {author} {\bibfnamefont
  {S.}~\bibnamefont {Stobbe}},\ and\ \bibinfo {author} {\bibfnamefont
  {O.}~\bibnamefont {Painter}},\ }\bibfield  {title} {\bibinfo {title} {A
  chip-scale integrated cavity-electro-optomechanics platform},\ }\href
  {https://doi.org/10.1364/OE.19.024905} {\bibfield  {journal} {\bibinfo
  {journal} {Opt. Express}\ }\textbf {\bibinfo {volume} {19}},\ \bibinfo
  {pages} {24905} (\bibinfo {year} {2011})}\BibitemShut {NoStop}%
\bibitem [{\citenamefont {Midolo}\ \emph {et~al.}(2012)\citenamefont {Midolo},
  \citenamefont {Yoon}, \citenamefont {Pagliano}, \citenamefont {Xia},
  \citenamefont {van Otten}, \citenamefont {Lermer}, \citenamefont
  {H\"{o}fling},\ and\ \citenamefont {Fiore}}]{2012OE}%
  \BibitemOpen
  \bibfield  {author} {\bibinfo {author} {\bibfnamefont {L.}~\bibnamefont
  {Midolo}}, \bibinfo {author} {\bibfnamefont {S.~N.}\ \bibnamefont {Yoon}},
  \bibinfo {author} {\bibfnamefont {F.}~\bibnamefont {Pagliano}}, \bibinfo
  {author} {\bibfnamefont {T.}~\bibnamefont {Xia}}, \bibinfo {author}
  {\bibfnamefont {F.~W.~M.}\ \bibnamefont {van Otten}}, \bibinfo {author}
  {\bibfnamefont {M.}~\bibnamefont {Lermer}}, \bibinfo {author} {\bibfnamefont
  {S.}~\bibnamefont {H\"{o}fling}},\ and\ \bibinfo {author} {\bibfnamefont
  {A.}~\bibnamefont {Fiore}},\ }\bibfield  {title} {\bibinfo {title}
  {Electromechanical tuning of vertically-coupled photonic crystal nanobeams},\
  }\href {https://doi.org/10.1364/OE.20.019255} {\bibfield  {journal} {\bibinfo
   {journal} {Opt. Express}\ }\textbf {\bibinfo {volume} {20}},\ \bibinfo
  {pages} {19255} (\bibinfo {year} {2012})}\BibitemShut {NoStop}%
\bibitem [{\citenamefont {Deotare}\ \emph {et~al.}(2012)\citenamefont
  {Deotare}, \citenamefont {Bulu}, \citenamefont {Frank}, \citenamefont {Quan},
  \citenamefont {Zhang}, \citenamefont {Ilic},\ and\ \citenamefont
  {Loncar}}]{2012NC}%
  \BibitemOpen
  \bibfield  {author} {\bibinfo {author} {\bibfnamefont {P.~B.}\ \bibnamefont
  {Deotare}}, \bibinfo {author} {\bibfnamefont {I.}~\bibnamefont {Bulu}},
  \bibinfo {author} {\bibfnamefont {I.~W.}\ \bibnamefont {Frank}}, \bibinfo
  {author} {\bibfnamefont {Q.}~\bibnamefont {Quan}}, \bibinfo {author}
  {\bibfnamefont {Y.}~\bibnamefont {Zhang}}, \bibinfo {author} {\bibfnamefont
  {R.}~\bibnamefont {Ilic}},\ and\ \bibinfo {author} {\bibfnamefont
  {M.}~\bibnamefont {Loncar}},\ }\bibfield  {title} {\bibinfo {title} {All
  optical reconfiguration of optomechanical filters},\ }\href
  {https://doi.org/10.1038/ncomms1830} {\bibfield  {journal} {\bibinfo
  {journal} {Nat. Commun.}\ }\textbf {\bibinfo {volume} {3}},\ \bibinfo {pages}
  {846} (\bibinfo {year} {2012})}\BibitemShut {NoStop}%
\bibitem [{\citenamefont {Iwase}\ \emph {et~al.}(2012)\citenamefont {Iwase},
  \citenamefont {Hui}, \citenamefont {Woolf}, \citenamefont {Rodriguez},
  \citenamefont {Johnson}, \citenamefont {Capasso},\ and\ \citenamefont
  {Lon{\v{c}}ar}}]{2012JMM}%
  \BibitemOpen
  \bibfield  {author} {\bibinfo {author} {\bibfnamefont {E.}~\bibnamefont
  {Iwase}}, \bibinfo {author} {\bibfnamefont {P.-C.}\ \bibnamefont {Hui}},
  \bibinfo {author} {\bibfnamefont {D.}~\bibnamefont {Woolf}}, \bibinfo
  {author} {\bibfnamefont {A.~W.}\ \bibnamefont {Rodriguez}}, \bibinfo {author}
  {\bibfnamefont {S.~G.}\ \bibnamefont {Johnson}}, \bibinfo {author}
  {\bibfnamefont {F.}~\bibnamefont {Capasso}},\ and\ \bibinfo {author}
  {\bibfnamefont {M.}~\bibnamefont {Lon{\v{c}}ar}},\ }\bibfield  {title}
  {\bibinfo {title} {Control of buckling in large micromembranes using
  engineered support structures},\ }\href
  {https://doi.org/10.1088/0960-1317/22/6/065028} {\bibfield  {journal}
  {\bibinfo  {journal} {J. Micromech. Microeng.}\ }\textbf {\bibinfo {volume}
  {22}},\ \bibinfo {pages} {065028} (\bibinfo {year} {2012})}\BibitemShut
  {NoStop}%
\bibitem [{\citenamefont {Espinel}\ \emph {et~al.}(2017)\citenamefont
  {Espinel}, \citenamefont {Santos}, \citenamefont {Luiz}, \citenamefont
  {Alegre},\ and\ \citenamefont {Wiederhecker}}]{2017SR}%
  \BibitemOpen
  \bibfield  {author} {\bibinfo {author} {\bibfnamefont {Y.~A.~V.}\
  \bibnamefont {Espinel}}, \bibinfo {author} {\bibfnamefont {F.~G.~S.}\
  \bibnamefont {Santos}}, \bibinfo {author} {\bibfnamefont {G.~O.}\
  \bibnamefont {Luiz}}, \bibinfo {author} {\bibfnamefont {T.~P.~M.}\
  \bibnamefont {Alegre}},\ and\ \bibinfo {author} {\bibfnamefont {G.~S.}\
  \bibnamefont {Wiederhecker}},\ }\bibfield  {title} {\bibinfo {title}
  {Brillouin optomechanics in coupled silicon microcavities},\ }\href
  {https://doi.org/10.1038/srep43423} {\bibfield  {journal} {\bibinfo
  {journal} {Sci. Rep.}\ }\textbf {\bibinfo {volume} {7}},\ \bibinfo {pages}
  {43423} (\bibinfo {year} {2017})}\BibitemShut {NoStop}%
\bibitem [{\citenamefont {Yang}\ \emph
  {et~al.}(2021{\natexlab{b}})\citenamefont {Yang}, \citenamefont
  {Albrow-Owen}, \citenamefont {Cai},\ and\ \citenamefont {Hasan}}]{2021S2}%
  \BibitemOpen
  \bibfield  {author} {\bibinfo {author} {\bibfnamefont {Z.}~\bibnamefont
  {Yang}}, \bibinfo {author} {\bibfnamefont {T.}~\bibnamefont {Albrow-Owen}},
  \bibinfo {author} {\bibfnamefont {W.}~\bibnamefont {Cai}},\ and\ \bibinfo
  {author} {\bibfnamefont {T.}~\bibnamefont {Hasan}},\ }\bibfield  {title}
  {\bibinfo {title} {Miniaturization of optical spectrometers},\ }\href
  {https://doi.org/10.1126/science.abe0722} {\bibfield  {journal} {\bibinfo
  {journal} {Science}\ }\textbf {\bibinfo {volume} {371}},\ \bibinfo {pages}
  {eabe0722} (\bibinfo {year} {2021}{\natexlab{b}})}\BibitemShut {NoStop}%
\end{thebibliography}%

\end{document}